\documentclass[published]{JHEP3} % 10pt is ignored!

\JHEP{00(2012)000}

\JHEPspecialurl{http://jhep.sissa.it/JOURNAL/JHEP3.tar.gz}

\usepackage{epsfig,multicol,bbm}

\newcommand\fverb{\setbox\fverbbox=\hbox\bgroup\verb}
\newcommand\fverbdo{\egroup\medskip\noindent%
			\fbox{\unhbox\fverbbox}\ }
\newcommand\fverbit{\egroup\item[\fbox{\unhbox\fverbbox}]}
\newbox\fverbbox

\title{Recovery of fluctuation spectrum evolution from tomographic
  shear spectra} 

\author{Silvio A. Bonometto$^{1,2,3}$ \& Marino Mezzetti$^{1,2}$, \\
  $^1$ -- Department of Physics, Astronomy Unit, Trieste University, Via
  Tiepolo 11, I~34143 Trieste, Italy \\ $^2$ -- I.N.A.F. --
  Astronomical Observatory of Trieste, Via Tiepolo 11, I~34143
  Trieste, Italy \\ $^3$ -- I.N.F.N. -- Sezione di Trieste, Via
  Valerio, 2 I~34127 Trieste, Italy \\ }

\received{.....}
\accepted{.....}
\abstract{Forthcoming large angle surveys are planned to obtain high
  precision tomographic shear data. In principle, they will allow us
  to recover the spectra of matter density fluctuation, at various
  redshift, through the inversion of the expressions yielding shear
  from fluctuation spectra. This was discussed in previous work, where
  $SVD$ techniques for matrix inversion were also shown to be the
  optimal tool to this aim. Here we show the significant improvements
  obtainable by using a 7 bin tomography, as allowed by future {\sc
    Euclid} data, and discuss error propagation from shear to
  fluctuation spectra. We find that the technique is a promising tool,
  namely for the analysis of baryon physics through high--$\ell$ shear
  spectra and to test the consistency between expansion rate and
  fluctuation growth.  }


\begin{document}

\section{Introduction}
Dark Energy (DE) is one of the main discoveries --$\, $and puzzles$\,
$-- in contemporary physics. Cosmic Microwave Background spectra or
Baryonic Acoustic Oscillations directly constrain its contribution to
the cosmic budget, but only simultaneous low--$z$ measures of cosmic
expansion rate and fluctuation growth can provide us real clues on DE
nature \cite{albrecht}. In fact, besides of its state equation $w(z)$,
we aim at knowing whether DE is a further physical dark component, or
its observations are a signal of new physics, as GR violations
\cite{mfg}, interactions between the dark components, possibly
suggesting a unified picture for them \cite{cde,bento}, or even more
exotic scenarios \cite{voids,back}. It is then worth outlining that,
while SNIa data constrain the expansion rate, tomographic shear data
enable us to follow the growth of density inhomegeneities.
% ellipse (on) 
Being directly sensitive to the whole mass distribution, they also
allow us to forget any problem related to light--mass conversion.
% ellipse (off) 

In order to exploit this kind of data, it has become customary to
follow a Bayesian approach. We then consider a set of parameters
spanning a wide set of models, for each of whom a number of observables
can be predicted; predictions are then compared with observational data
and relative errorbars, so obtaining likelihood hyper-elipso\"\i ds in
the parameter space. Similarly, when cosmic shear data will be
available, one shall predict the tomographic shear spectra
$C_{ij}(\ell)$ (the indeces $i,j$ run on the tomographic bins) and
enrich the fit to data by comparing them with tomographic shear data.

In a previous paper (\cite{PaperI}, Paper I herebelow) we considered a
different option: directly deriving the spectra $P(k,z)$ at various
redshift $z$ from $C_{ij}(\ell)$.  A prediction of shear spectra is
then no longer required, and shear data can soon be compared with
other data on fluctuation growth, to be possibly used in parallel with
them. To do so, we need to invert the integral relation yielding
$C_{ij}(\ell)$ from $P(k,z)$. As we shall see, this option makes
sense, in particular, if we wish to inspect background geometry and
fluctuation dynamics separately.

In Paper I, our analysis was restricted to 5 tomographic bins. On the
contrary, {\sc Euclid}\footnote{www.euclid-ec.org} data, obtained
from a celestial area of 15,000 square degrees with a median redshift
$z_m = 0.9$, will allow N--bin tomography, with N$\gg 5$ and up to 10
\cite{laureijs}.

In this paper, we consider a 7 bin tomography and find that the
improvements allowed by this N increase, when aiming to recover
$P(k,z)$, are substantial. {\bf We shall also briefly comment on the
  reasons why attempting spectral recovery with $N>7$ leads to
  problems.}  The recovery still uses the singular value decomposition
(SVD) technique for matrix inversion, an approach already followed by
different authors in various physical contexts \cite{svdauth,ez}.
Specific options of this technique, suitable to treat quasi--singular
matrices, will be however used here for the first time in this
context. Furthermore, we shall treat the question of noise propagation
from shear to fluctuation spectra. Matrix theory allows us to predict
upper limits to error magnification in such transition. Lukily enough,
in this specific case, such limits are only marginally approached,
while a simple filter allows us further noise reduction.

Let us also notice that, besides of verifying the coherence between
expansion and fluctuation growths, this inversion technique could also
facilitate the discrimination between different options on baryon
physics, shaping $P(k,z)$ at high--$k$.
% ellipse (correction above) 

In order to test the inversion algorithm we need to input fluctuation
spectra for a given model. We use them to build fluctuation spectra
$P(k,z)$ at various redshifts and, from them, the shear spectra
$C_{ij}(\ell)$. Successively, starting from $C_{ij}(\ell),$ we test
how efficiently $P(k,z)$ is recovered. Being also interested in baryon
physics and, therefore, having~to~exp\-lore the high--$\ell$ region,
recent hydrodynamical simulations are used and briefly discussed. Let
us however remind that the lensing spectra can be roughly shared in 3
$\ell$--ranges (see Figure \ref{kl1a}, here below): The $C_{ij}(\ell)$
for $\ell < \sim 500$ essentially feel just the linear dynamics, even
for $i,j = 1$. For $500 < \ell < 1500$ the contribution coming from
{\it non--linear $k$'s} becomes relevant. At $\ell >2000$, shear data
start to exhibit a dependence on baryon physics.

The plan of the paper is as follows: In the next Section we shall
review the procedure allowing to pass from data on galaxy ellipticity
to shear spectra. In particular, we shall obtain the window functions
in the 7--bin case, taking into account that, for most lensed
galaxies, only photometric redshift will be avaiable.

In Section 3 fluctuation and shear spectra are defined and their
relation set in an operational form. Section 4 debates the procedure
of Gauss--Laguerre integration to obtain shear spectra from
fluctuation spectra. Section 5 introduces the formal procedure to
invert the equation yielding $C_{ij}(\ell)$ from $P(k,z)$, also based
on Gauss--Laguerre integration. Section 6 then introduces the SVD
technique and uses it for the study of the singularity level, as well
to recover $P(k,z)$ from $C_{ij}(\ell)$, in 3 cases: (i) When
$C_{ij}(\ell)$ is formally obtained with a GL integration
technique. (ii) When $C_{ij}(\ell)$ is ``exact''. (iii) When noise is
added to it. Results are summarized and discussed in Section 7 and
conclusions are drawn in Section 8, where we also outline possible
options to improve the present approach.

\section{Evaluation of shear--shear correlation functions}
Using measured galaxy ellipticities
\cite{Kamionkowski1998,Crittenden2008,Stebbins} to gauge gravitational
shear is a non--trivial task. Uncorrelated random ellipticities
scarsely matter, the problem being {\it intrinsic} shear, due to
nearby galaxy alignemnt. Assuming that data can reliably cleansed from
it may be premature, but here we let apart this problem and proceed as
though ellipticities due to gravitational shear only are measured (see
however \cite{BK,sch,JB}, and references therein).

% ------- ellipse (off)

We shall also assume (spatially flat) models whose background metric
reads
\begin{equation}
ds^2 = a^2(\tau)[d\tau - d\lambda^2]~,
\label{metric}
\end{equation}
so that $\tau$ is the conformal time, $d\lambda$ being the comoving
distance element, and $a(\tau) \equiv (1+z)^{-1}$ the scale factor.
If $\tau_0$ is the present time, 
\begin{equation}
u(z) = \tau_0-\tau(z)
\end{equation}
is the conformal time distance --as well as the comoving distance--
from $z$.

Let us also assume that galaxies observed in unit solid angle have a
redshift distribution
\begin{equation}
n(z) = {d^2 N \over d\Omega\, dz} = {\cal C} ~\bigg({z \over z_0
}\bigg)^A \exp\bigg[- \left( z \over z_0 \right)^B \bigg]
\end{equation}
with  $A=2$, $B=1.5$ so that
\begin{equation}
{\cal C} = {B / \left[z_0 \Gamma \left( A+1 \over B
\right) \right]}  = {1.5 / z_0}
\label{nz}
\end{equation}
(here $z_0 = z_m/1.412$ while the median redshift $z_m = 0.9~$, in
agreement with {\sc Euclid} specifications \cite{laureijs}). These
phaenomenological values can be suitably modified, if needed.

To appreciate the effects of fluctuation evolution, galaxies are then
shared into $N$ redshift bins, with limits $z_r$ selected so that they
contain equal numbers of galaxies. For large galaxy sets, photometric
redshifts only will be available and, to evaluate the expected
distribution on redshift for the $r$--th bin galaxies, we can apply
the filter
\begin{equation}
\Pi_r (z) = \int_{z_{r}}^{z_{r+1}} dz' ~ {e^{-{(z - z')^2 \over 2 \sigma^2 (z)} }
 \over \sqrt{2 \pi}~ \sigma(z)}  ~=~
%\sigma(z)} \exp \left[-{(z - z')^2 \over 2 \sigma^2 (z)} \right] =
 {1 \over 2}
\left[
{\rm Erf}\left(z_{r+1}-z \over \sqrt{2}\sigma(z)
\right)-{\rm Erf}\left[z_{r}-z \over \sqrt{2}\sigma(z)\right]
\right]
\end{equation}
% ellipse: _{ph} canceled 
to $n(z)$, as suggested, e.g., in \cite{a15-16}. In this way we obtain
the distribution
\begin{equation}
D_r(z) = n(z) \Pi_r(z)
\label{diz}
\end{equation}
whose integrals are $\simeq 1/N$, their sum being (exactly) $n(z)$.
In this work we shall take $\sigma(z) = 0.05~(1+z)$, coherently with
{\sc Euclid} expectations \cite{laureijs} (see also
\cite{amaraAmendola}).

%%%%%%%%%%%%%%%%%%%%%%%%%%%%%%%%%%%%%%%%%%%%%%%%%%%%%%%%%%%%%%%%%%%%%%%
\begin{figure}
\begin{center}
\vskip -.9truecm
\includegraphics[height=7.3cm,angle=0]{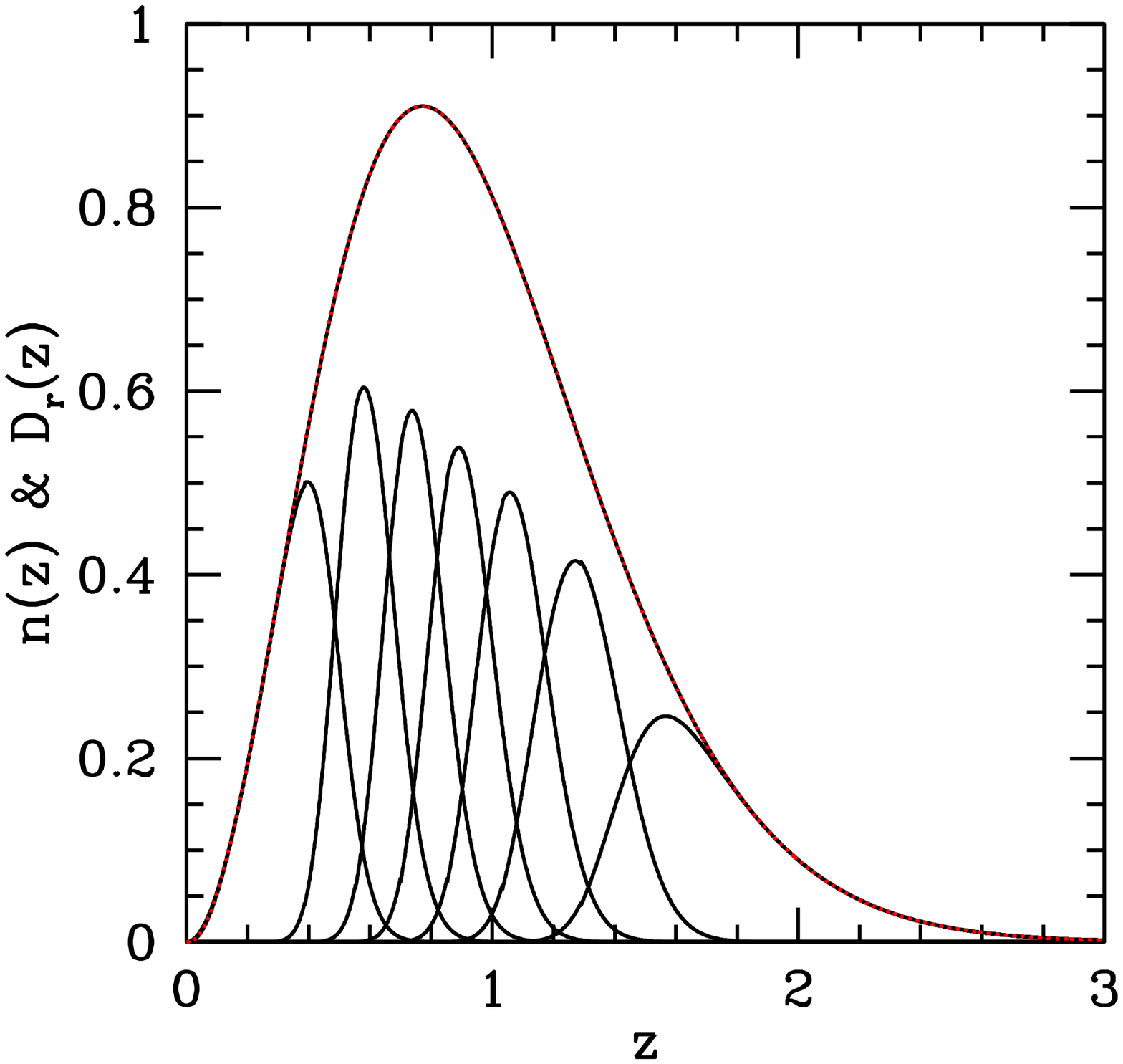}
\includegraphics[height=7.3cm,angle=0]{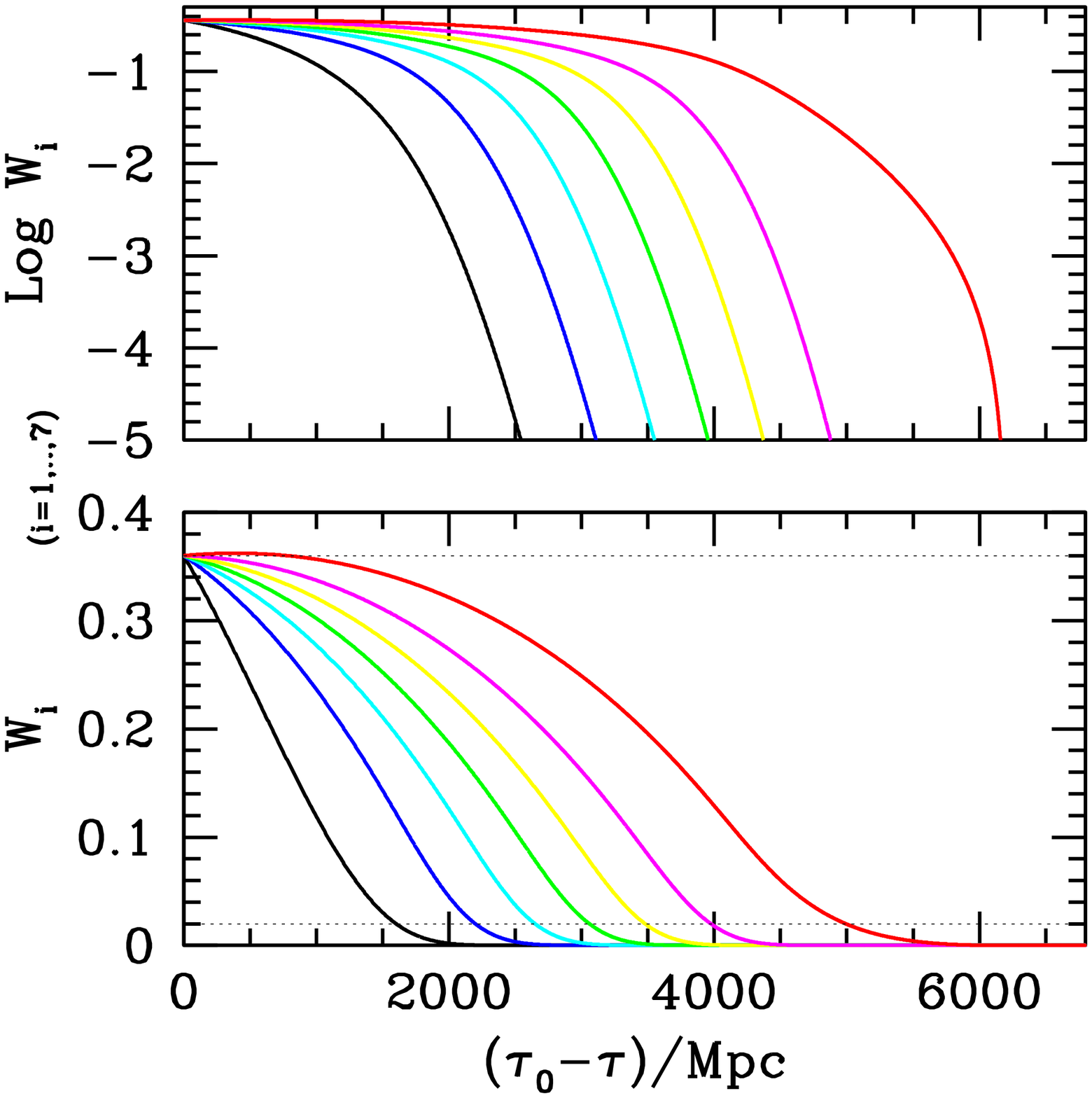}
\end{center}
\vskip -.8truecm
\caption{{\it l.h.s.:} Total galaxy distribution $n(z)$ (red curve) and
  distributions $D_r(z)$ of the galaxies in $N=7$ bins as obtainable
  though photometric redshift values. Their sum (black dots) overlaps
  $n(z)$. {\it r.h.s.:} Window functions for the same 7 bin case. }
\label{dW}
\end{figure}
%%%%%%%%%%%%%%%%%%%%%%%%%%%%%%%%%%%%%%%%%%%%%%%%%%%%%%%%%%%%%%%%%%%%%%%
The images of galaxies belonging to a bin are lensed by the gravity of
matter laying at lower $z$. This is taken into account by defining the
{\it window functions}
\begin{equation}
\label{WI}
W_r (z) = {3 \over 2} \Omega_m (1+z) \int_{\Delta z_r}
dz'~\delta_r(z')~ {\cal P}\left[u(z')-u(z) \over u(z') \right]~;
\end{equation}
here integration is restricted to the intervals $\Delta z_r$, where
$D_r$ is non--zero, while
\begin{equation}
\delta_{r}(z) = {D_{r}(z) / \int_{0}^{\infty}D_{r}(z')dz'}
\end{equation} 
and $\cal P$$(x) = x$ or~0 if $x>0$ or $<0$: only systems closer than
a galaxy can distort its image. In Figure~\ref{dW} we show $n(z)$ and
$D(z)$ (l.h.s.), as well as $W_r(u)$ (r.h.s.), for the 7--bin case.

The cosmology used in this paper is the same of Paper I: a
$\Lambda$CDM model with density parameters for matter and baryons,
Hubble parameter, primordial spectral index and m.s. amplitude
of density fluctuations, at $z=0$, on the scale of 8$\, h^{-1}$Mpc,
shown herebelow

\vskip .1truecm
\centerline{{\rm Table I}} \vglue -.1truecm \centerline{
  ----------------------------------------------------------------- }
\vglue -.5truecm
$$
\matrix{
\Omega_m  & \Omega_b       & h    & n_s & \sigma_8 \cr
0.24 & 4.13 \times 10^{-2} & 0.73 &  0.96 & 0.8
}
$$ 
\vglue -.1truecm
\centerline{ ----------------------------------------------------------------- }

\vskip -.02truecm
\noindent
To build the functions $W_r(z)$ we need to know $u(z)$.  However, (i)
no baryon physics is needed, not even the $\Omega_b$ value; (ii)
fluctuation parameters, as $\sigma_8$ or $n_s$, are also not needed.
This is to be borne in mind in view of the final discussion.

\section{Density fluctuation and shear spectra}
%%%%%%%%%%%%%%%%%%%%%%%%%%%%%%%%%%%%%%%%%%%%%%%%%%%%%%%%%%%%%%%%%%%%%%%
\begin{figure}
\begin{center}
\vskip -4.2truecm
\includegraphics[height=11.2cm,angle=0]{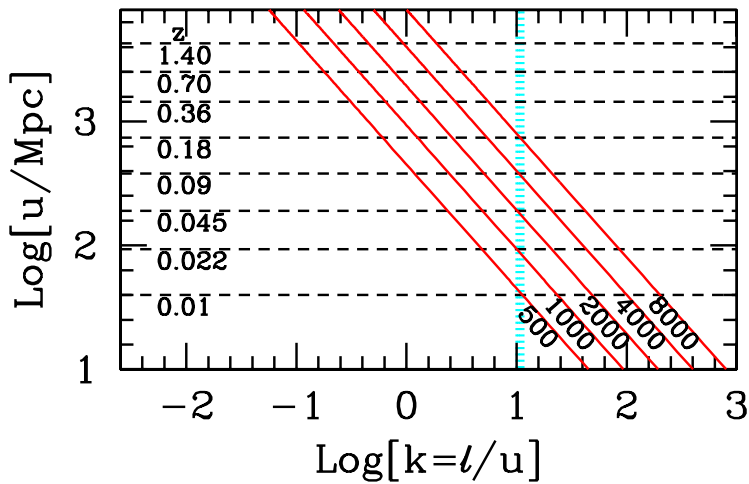}
\end{center}
\vskip -1.truecm
\caption{Integration patterns: red lines show $k$ values contributing
  to $C_{ij}(\ell)$ at given $\ell$ (aside them). At the r.h.s. of the
  cyan band, baryon physics yield significant contributions to
  spectra.  Its slight (model dependent) decrease with $z$ is
  neglected here.  }
\label{kl1a}
\vskip -.4truecm
\end{figure}
%%%%%%%%%%%%%%%%%%%%%%%%%%%%%%%%%%%%%%%%%%%%%%%%%%%%%%%%%%%%%%%%%%%%%%%
The shear--shear correlation functions $\xi_{+/-}^{ij}(\theta)$,
derived from the distributions of the ellipticities of galaxies in the
$i,j$--th bands, can be expanded in spherical harmonics yielding
\begin{equation}
\label{discrete}
\xi_{+/-}^{ij}(\theta) =
(2\pi)^{-1} \int_0^\infty d\ell\, \ell\, J_{0/4}(\ell \theta)\, C_{ij}(\ell)
\end{equation}
$J_n$ being first kind Bessel functions of order $n$. In turn, the shear
spectra
\begin{equation}
C_{ij}(\ell) = H_{0}^{~4} \int_{0}^{\tau_0 } du~ W_{i}(u) W_{j}(u)~ P[\ell/u,u]~;
\label{pijl}
\end{equation}
are related to the power spectrum $P(k,z) = \langle| \delta
(k,z)|^2\rangle$; here $\delta(k,z)$ is the Fourier trasform of the
density fluctuation field at redshift $z$. The functions $W_i(u)$
weight the contributions to shear correlations on the angular scale
$\theta (\sim 2\pi/\ell)$, from density fluctuations over linear
scales $\lambda (\sim 2\pi/k=2\pi u/\ell)$, increasing with $z$.
Accordingly, shear correlations at a given $\ell$ receive
contributions from decreasing $k$ values as $u$ increases. In turn,
for $u \to 0$, $P(k,u)$ should be evaluated at $k \to \infty$ where,
however, it vanishes. Figure \ref{kl1a} shows typical integration
patterns on the $\log k$--$\log u$ plane, amounting to tilted straight
lines.

\section{Gauss--Laguerre integration procedure}
Our aim now amounts to finding a technique to invert eq.~(\ref{pijl})
and to testing such technique; accordingly, eq.~(\ref{pijl}) is used
first to work out the spectra $C_{ij}(\ell)$ for the given model. Then
we test the algorithm --that we shall build-- to reobtain $P(k,z)$
from $C_{ij}(\ell)$.

Fluctuation spectra at $k > \sim 10\, h$Mpc$^{-1}$ depend
significantly on baryon physics. As substantial contributions to the
integral (\ref{pijl}) arise for $z >\sim 0.01$, Figure \ref{kl1a} shows
that only $\ell \sim 500$ is (almost) free of baryon physics while, at
$\ell >\sim 1500$, it becomes critical. Accordingly, to work out
sensible $C_{ij}(\ell)$ spectra for $\ell > 1500$, we need $P(k,z)$
obtained from hydro simulations. Analytical approximated expressions
as {\sc Halofit} are useful just for tests. In Appendix B we discuss
the simulation used (the same as in Paper I) and interpolatory
techniques. Here we assume $P(k,z)$ to be available at any $k$ and $z$
value.

Besides of performing the integral (\ref{pijl}) with an ``exact''
adaptive Riemann integration, we need to consider a Gaussian
integration procedure as well. It amounts to projecting each integrand
function $f(u)= W_{i}(u) W_{j}(u) P(l/u,u) $ onto polynomials
$\pi_\alpha(u)$, orthogonal with an assigned weight function $R$$(u)$,
finding its components $f_\alpha$. As the integrals $\Pi_\alpha$ of
each polynomial are known, $f(u)$ integral reads $\sum_{\alpha=1}^\nu
f_\alpha \Pi_\alpha$. This is a reliable value if $\nu$ is large
enough. This technique can be translated into a simple procedure,
i.e. to a weighted sum of values taken by the integrand function
$f(u)$ in a suitable set of points~$u_\alpha$. 

More in detail, by using {\it monic} polynomials (obtained from a
recurrence relation assuming that the coefficient of the leading term,
for each $\alpha$, is unity) we have that
\vskip -.3truecm
\begin{equation}
  \int_0^\infty dx \, \, R(x) \, \pi_\alpha(x) \pi_\beta(x) = {\cal N}
  \delta_{\alpha\beta} \, \, ,
\end{equation}
\vskip -.1truecm
\noindent
with a known normalization $\cal N$. The $\nu$ zero's of $\pi_\nu (x)$
are the points $x_\alpha$, while
\vskip -.3truecm
\begin{equation}
w_\alpha = {\int_0^\infty dx\, \, R(x) \pi_{\nu-1}^2(x)
\over \pi_{\nu-1}(x_\alpha) \pi'_{\nu}(x_\alpha)} \, \, \, ,
\label{wr}
\end{equation}
are the corresponding weights ($\pi' (x)$ being the ordinary
derivative of $\pi (x)$). Accordingly
\begin{equation}
\int_0^\infty du\, f(u) = \int_0^\infty dx\, F(x)\, R(x) = \sum_\alpha
w_\alpha F(x_\alpha)
\label{sum}
~~~~{\rm with}~~~~
F(x) = {f[u(x)]\over R(x)}{du \over dx}~.
\end{equation}
For $R(x) \propto e^{-x}$, the technique is dubbed Gauss--Laguerre
(GL) integration, as $\pi_\alpha (x) = L_\alpha (x)$, the Laguerre
polynomials. The integral (\ref{pijl}) is cut off by the decay of the
$W_i$~func\-tions and GL integration can be applied by taking $x =
(u/\bar u)^\beta$, suitably selecting $\bar u$ and~$\beta$.

Let us then rewrite eq. (\ref{pijl}) as follows:
\vskip -.5truecm
\begin{equation}
\label{linear1}
c_A(\ell) = \sum_{r=1}^\nu w_r S_{A x_r} p_{x_r} (\ell) \equiv
\sum_{r=1}^\nu {\cal M}_{Ar} p_{r} (\ell)\, \, \, ~~{\rm being }~~ c_A =
C_{ij}/H_0^4~,~~ A \equiv ij
\end{equation}
with 
\begin{equation}
S_{A x_r} = W_i(x_r) W_j(x_r) u_r/[\beta x_rR(x_r)]\, ,
\end{equation}
\begin{equation}
p_{r} (\ell) \equiv p_{x_r} (\ell)= 
P_\ell(u_r) = P(\ell/u_r,u_r) \,  ,
\end{equation}
being $u_r = u(x_r)$, and
the correspondence law
$$
\matrix{
i,j & 1,1 & ... & 1,7 & 2,2 & ... & 2,7 & 3,3 & ...& ...  & 7,7 \cr
A & 1 & ... & 7 & 8 & ... & 13 & 14 &  ... & ... & 28
}
$$

\section{Formal inversion}
If, in eq. (\ref{linear1}), we take $\nu=28$, $ {\cal M}_{Ar} $ is a
square matrix and, provided that it is non--singular, the inverse
equation
\begin{equation}
 p_{x_r} (\ell) = \sum_A ({\cal M})^{-1}_{rA} c_A(\ell)
\label{solution}
\end{equation}
also holds. In principle, we should then be able to work out the
spectrum $P(k,z)$ for any $k = \ell/u_r$, at the redshift values $z_r
= z(u_r) $. The inversion procedure acts on each $\ell$ value
separately and we could even perform a different choice of $\bar u$
and $\beta$ for each $\ell$.

In Paper I, $\nu$ could reach atmost 15. This led to 2 difficulties:
(i) 15 gaussian points could not allow us a nearly--exact integration;
(ii) the determinant of the $15 \times 15$ $\cal M$ matrix was close
to singular. An inversion could however be achieved by passing from 15
to 12 integration point. Although worsening the (i) point, this
allowed us a redundant system of equations, which could be treated
through the SVD technique. A suitable choice of $\bar x$ and $\beta$
then allowed us to overcome the (ii) point and to approach an exact
inversion of the $C_{ij}(\ell)$ spectra formally obtained through
Gaussian integration.

However, even this was not enough when trying to invert the results of
exact Riemann integration: the recovered $P(k,z)$ were systematically
ill--normalized. Besides of the cosmological parameters needed to
build the $W_r(u)$, we then needed to add $\Omega_b$ and $\sigma_8$ to
predict the normalization (the dependence on $n_s$ is negligible). In
this way we obtained all $P(k)$ details of at high--$k$ from
$C_{ij}(\ell)$, but the linear $P(k,z)$ was assumed to be known.

Here we aim at testing whether 28 indipendent equations can do better. In
particular, we aim at avoiding any input concerning baryonic and/or
fluctuation parameters.

\section{The SVD technique}
The risk of singularity for the matrix
\begin{equation}
{\cal M}_{Ar} = w_r W_{i(A)}(x_r) W_{j(A)}(x_r) u_r/[\beta x_rR(x_r)]\
\end{equation}
arises from the vanishing of low--$i$ window functions $W_i$, at
distances $u_r$ (and thence at the corresponding $x_r$ values) where
high--$i$ window functions are still significant. 

The SVD technique, often used in science to handle complex datasets
\cite{svdauth}, here is used to evaluate, first of all, the degree of
singularity of ${\cal M}_{Ar}$. The technique is based on a powerful
theorem of linear algebra, stating that any real $N_r \otimes N_c$
matrix $\cal M$, with $N_r \geq N_c$, can be decomposed, in a unique
way apart trivial overall factors, into a rows $\times$ columns
product
\vskip -.3truecm
\begin{equation}
{\cal M}_{N_r,N_c} = {\cal U}_{N_r,N_c} 
 \times
\left|~diag(s_i)~\right|_{N_c}  \times {\cal V}_{N_c,N_c}^T
\end{equation}
Here, in each index site, we set the range allowed to the index there.
Both $\cal M$ and $\cal V$ ($\cal V$$^T$ is its transposed) are
orthonormal matrices; also $\left|~diag(s_i)~\right|_{N_c} $ is a
(fully diagonal) matrix.

If $N_r = N_c$, the inverse of $\cal M$, in general, reads
\begin{equation}
{\cal M}^{-1} = {\cal V} \times \left|~diag(1/s_i)~\right| \times {\cal
  U}^T~,
\end{equation}
and the technique does not only invert large matrices, but also
handles them when inversion is numerically hard.  Fair manuals
describing the SVD technique are \cite{NR, golub}.

The $s_i$ components tell us the singularity level. When some of them
vanishes, the matrix is singular and the problem is said to be {\it
  ill--conditioned}. Even if it is not so, as in our cases, the ratio
between maximum and minimum $s_i$ (condition number) can exceed $\sim
10^6$ ($10^{12}$), and then there is no way to invert ${\cal M}$ in
{\it single} ({\it double}) precision. Aiming at a precision $\cal
O$$(1:10^5$--$10^6)$, the top $s_i/s_j$ ratio must be within $\sim
10^7$. This is a necessary condition, however, not a sufficient one
and we often need to be more restrictive.

\subsection{Singularity level}
\begin{figure}
\begin{center}
\vskip -3.5truecm
\includegraphics[height=9.cm,angle=0]{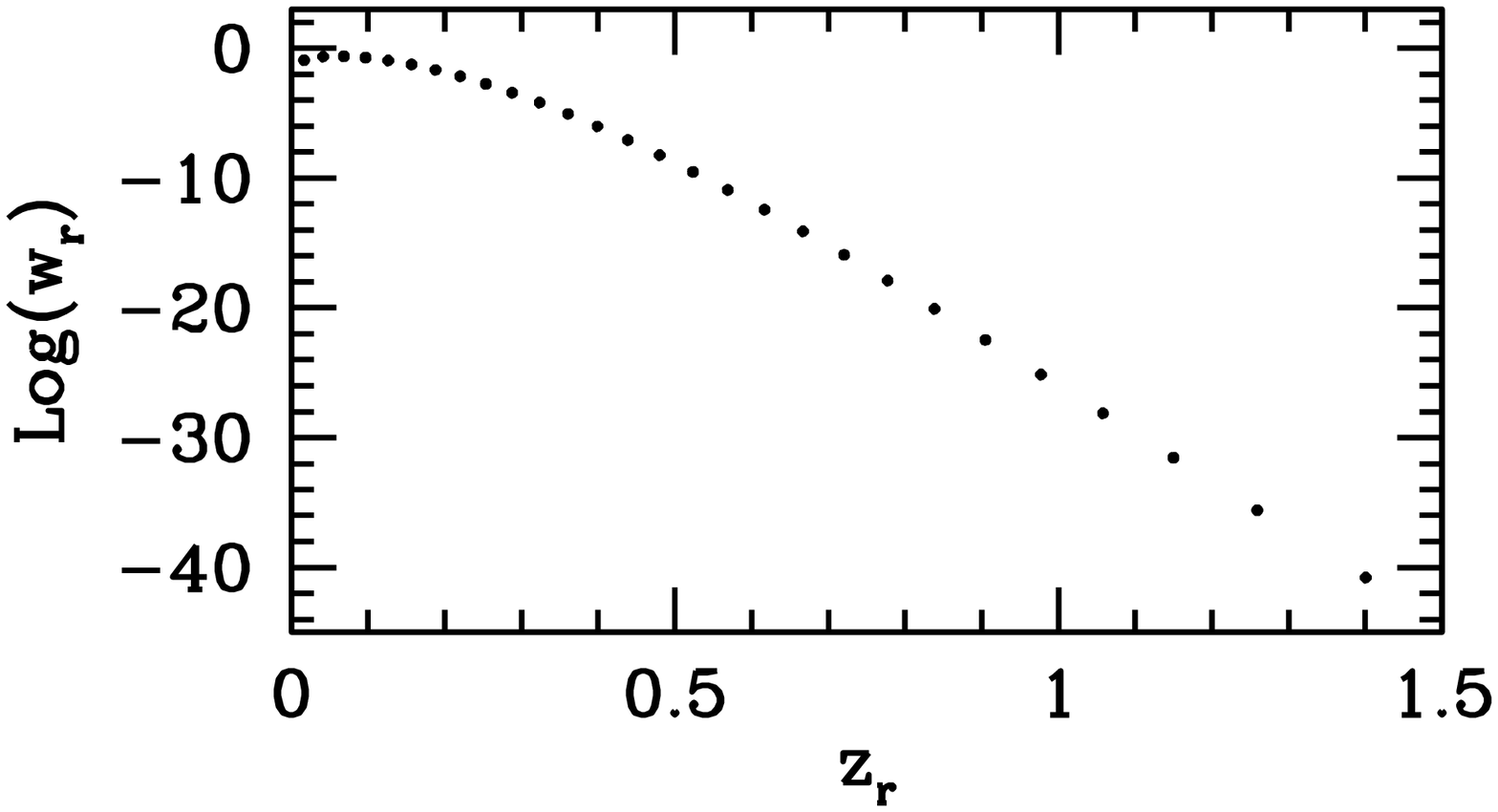}
\includegraphics[height=5.5cm,angle=0]{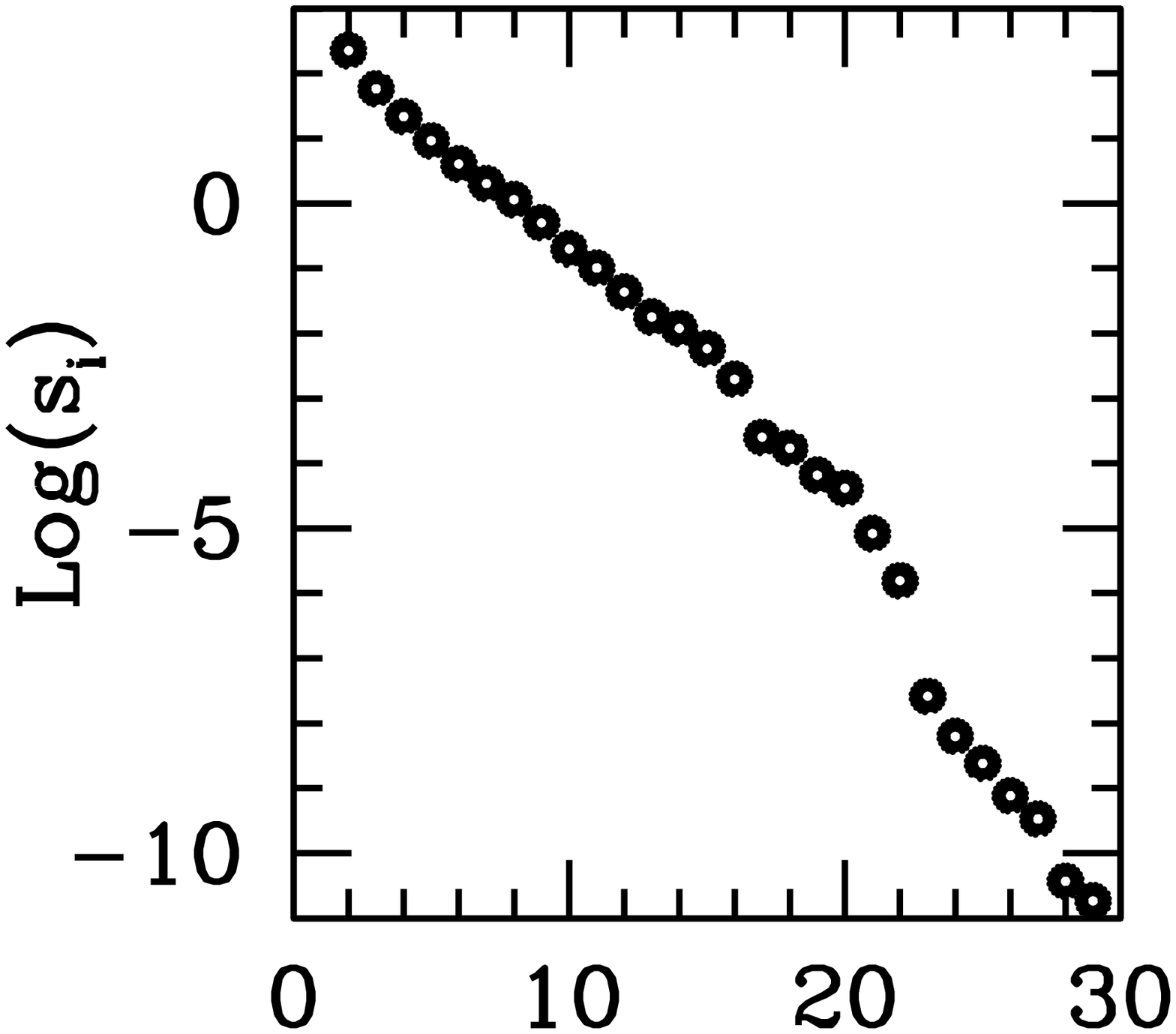}
\end{center}
\vskip -1.truecm
\caption{{\it L.h.s.:} Gauss--Laguerre weights vs.~the related
  redshift values.  {\it R.h.s.:} The elements $s_i$ of the diagonal
  matrix, yielding the singularity level of the matrix to be inverted.
}
\label{gaulag}
\end{figure}

Our first aim should amount to inverting the full 28$^2$ matrix. Its
condition number depends on the choice of $\bar u$ and $\beta$. Best
results are obtainable, as we discuss in Appendix B, when the top
$z_r\sim 1.40$ and the low--$z$ domain is sampled at $z >\sim
0.02$~. In Figure \ref{gaulag} the Gaussian weights vs.~the related
redshift values and the components of the diagonal matrix $s$ are
shown for $\bar u = 351$, $\beta = 1.8424$.
%%%%%%%%%%%%%%%%%%%%%%%%%%%%%%%%%%%%%%%%%%%%%%%%%%%%%%%%%%%%%%%%%%%%%%%
%%%%%%%%%%%%%%%%%%%%%%%%%%%%%%%%%%%%%%%%%%%%%%%%%%%%%%%%%%%%%%%%%%%%%%%
%%%%%%%%%%%%%%%%%%%%%%%%%%%%%%%%%%%%%%%%%%%%%%%%%%%%%%%%%%%%%%%%%%%%%%%
\begin{figure}
\begin{center}
\vskip .01truecm
\includegraphics[height=6.3cm,angle=0]{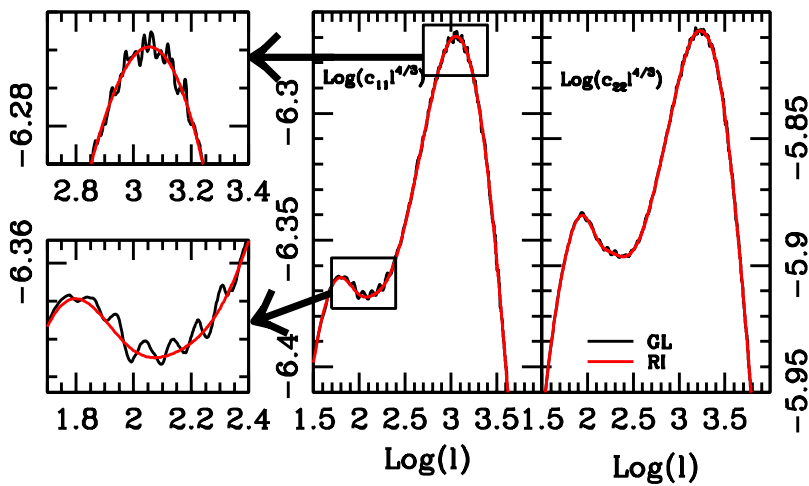}
\end{center}
\vskip -.6truecm
\caption{GL and RI integration compared. The magnified frames at the
  l.h.s. outline residual oscillations in GL integrals. }
\label{spettri}
\end{figure}
%%%%%%%%%%%%%%%%%%%%%%%%%%%%%%%%%%%%%%%%%%%%%%%%%%%%%%%%%%%%%%%%%%%%%%%
Each $w_r$ is multiplied by $e^{x_r}$, so that also $w_r \sim
10^{-40}$ are not negligible. On the contrary, the $s_i$ components do
span $\sim 13$~o.o.m. and inverting the $\cal M$ matrix, as it is,
requires unusual computational resources.
  
The SVD approach however tells us how to deal with such a case,
keeping as much information as possible. We can do so by simply
setting to zero the unwanted $s_i$ values. For instance, once the last
7 (8) $s_i$ are zeroed, the residual $s_i$ just span 8 (7) orders of
magnitude. This reduction, as explained in the cited manuals and
discussed in Appendix B, is quite different from disregarding a number
of equations: power spectra are recovered at all 28 redshifts and
discrepancies from input spectra are safely small for $0.1 \sim z \sim
1.2$. As a matter of fact, this is the very range where shear signals
are reliable, and the matrix inversion technique substantially yields
a best--fit recovery of input spectra.

Data on $C_{ij}(\ell)$ spectra will however come with errors and this
is a critical point to apply this approach. To study their impact, we
add a random Gaussian noise to the $C_{ij}(\ell)$ spectra worked out
from $P(k,z)$. Inversion bursts the noise: an upper limit to noise
magnification is set by the ratio between non--zero top and bottom
$s_i$ (see Appendix B).

For instance, if such $s_i$ span 8 orders of magnitude, we are likely
to obtain $C_{ij}(\ell)$~comple\-tely covered by noise. We therefore
need to exploit the SVD technique at its limits, by keeping just a
minimum number of non--zero $s_i$. Lukely enough, the transfered noise
is then significantly below the theoretical upper limit and can be
furtherly reduced by filtering.

In the next subsections we shall provide quantitative details on
these points.

\subsection{Recovery in the absence of noise}
Two different integration technique have been considered here: (i) an
adaptive Riemannian technique (RI, hereafter) allowing any wanted
precision; (ii) a Gauss--Laguerre technique (GL, hereafter), that
opens the door to inversion.

Although using 28 points and coefficients, GL does not allow us a full
recovery of the $C_{ij}(\ell)$ obtained through RI. Although smoothed,
simulation spectra keep some irregularities even for small $k$
increments. Even a 28--th degree polinomial is unable to follow all of
them.

In Figure \ref{spettri} we compare the results of GL and RI
integration for $C_{11}(\ell)$ and $C_{22}(\ell)$ (the coefficient
$H^4$ is omitted). The tiny differences ($\cal O$$.1 \, \%$ at most)
can be perceived, namely at low--$\ell$, where GL integrals exhibit
tiny oscillations. When GL integration uses more usual numbers of
points ($\sim 10$), these oscillations risk to be confused with BAO
effects. A similar danger exhists if RI integration follows an
interpolation of type (i), with not enough spectra along $z$. Spurious
BAO's are then an effect of the interpolating polinomials and
disappear when we increase the number of spectra along~$z~.$

\begin{figure}
\begin{center}
\vskip -2.2truecm
\includegraphics[height=6.9cm,angle=0]{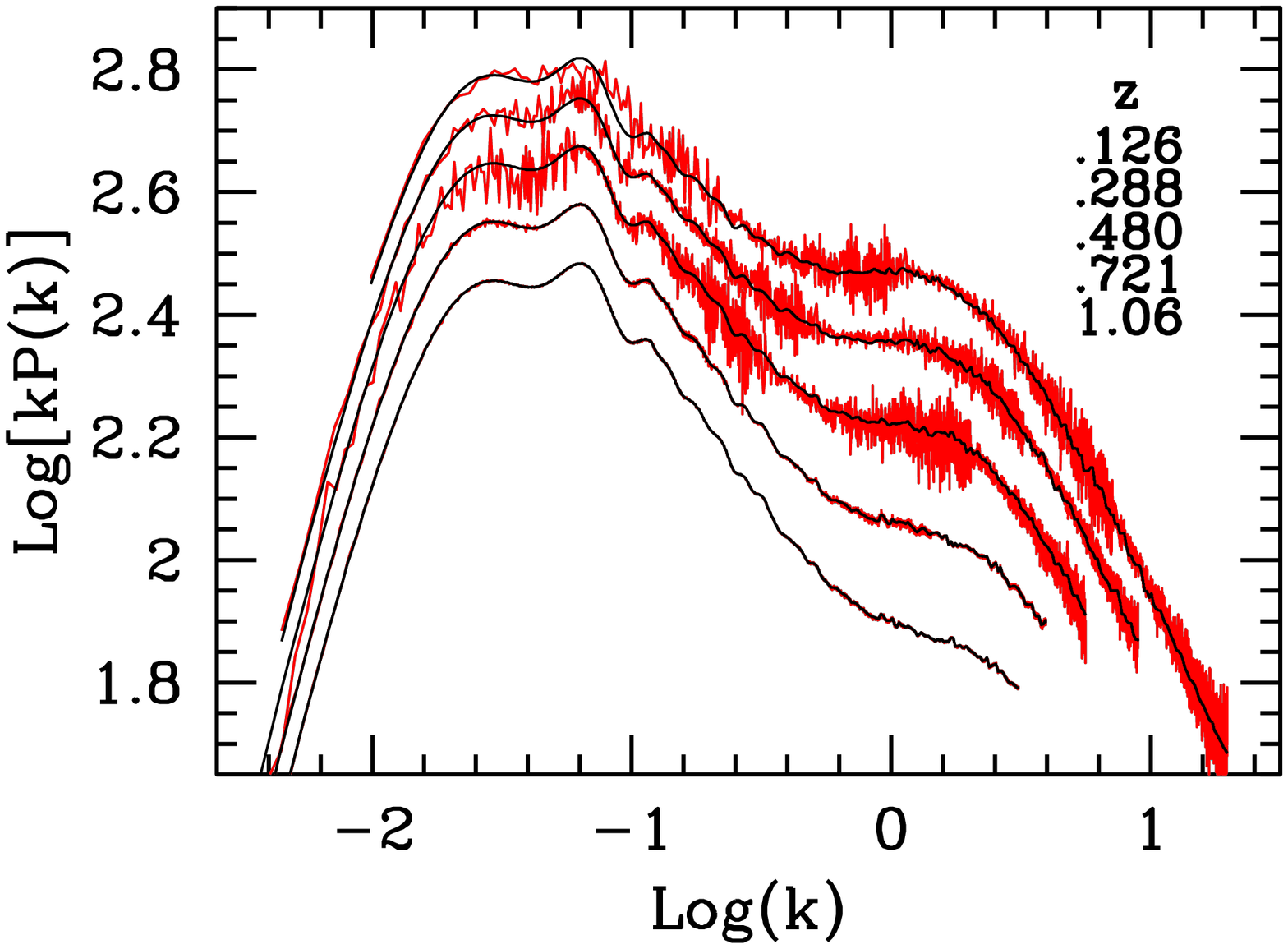}
%\vskip -2.truecm
\includegraphics[height=6.9cm,angle=0]{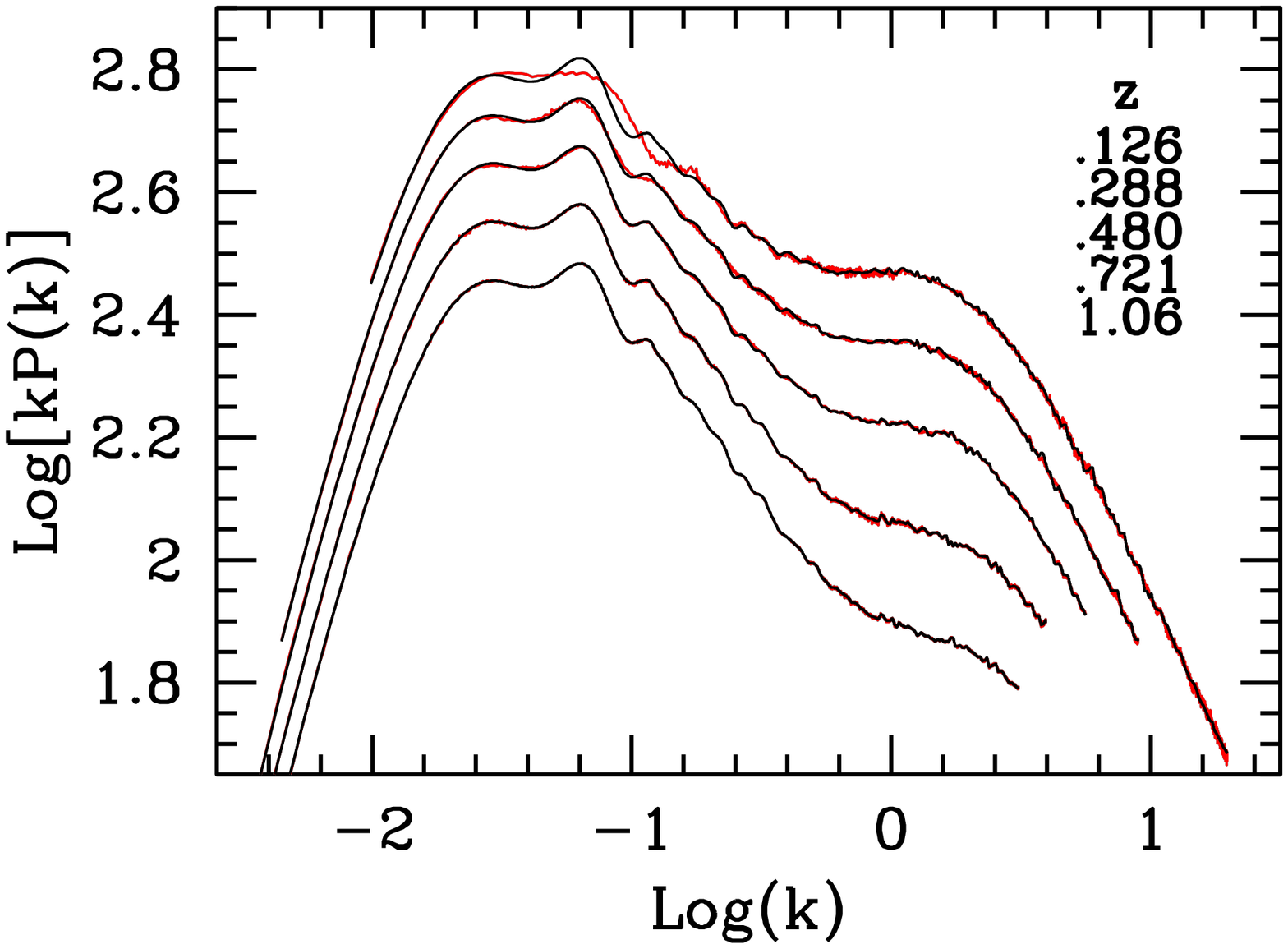}
\end{center}
\vskip -.6truecm
\caption{Recovery of $P(k,z)$ spectra, at the 5 redshifts shown in the
  frames, by using GL shear spectra and 21 (l.h.s.) or 20 (r.h.s.)
  non--zero $s_i$. Black (red) curves are the input (recovered)
  spectra.  In the r.h.s. plot, recovery is so efficient that red
  lines mostly desappear below black ones. }
\label{SPGL2021}
\end{figure}

In Figure \ref{SPGL2021} we show the results of matrix inversion
if applied to GL spectra, using 20 or 21 $s_i$ values.
%%%%%%%%%%%%%%%%%%%%%%%%%%%%%%%%%%%%%%%%%%%%%%%%%%%%%%%%%%%%%%%%%%%%%%%
%%%%%%%%%%%%%%%%%%%%%%%%%%%%%%%%%%%%%%%%%%%%%%%%%%%%%%%%%%%%%%%%%%%%%%%
%%%%%%%%%%%%%%%%%%%%%%%%%%%%%%%%%%%%%%%%%%%%%%%%%%%%%%%%%%%%%%%%%%%%%%%
\begin{figure}
\begin{center}
\vskip -2.2truecm
\includegraphics[height=6.9cm,angle=0]{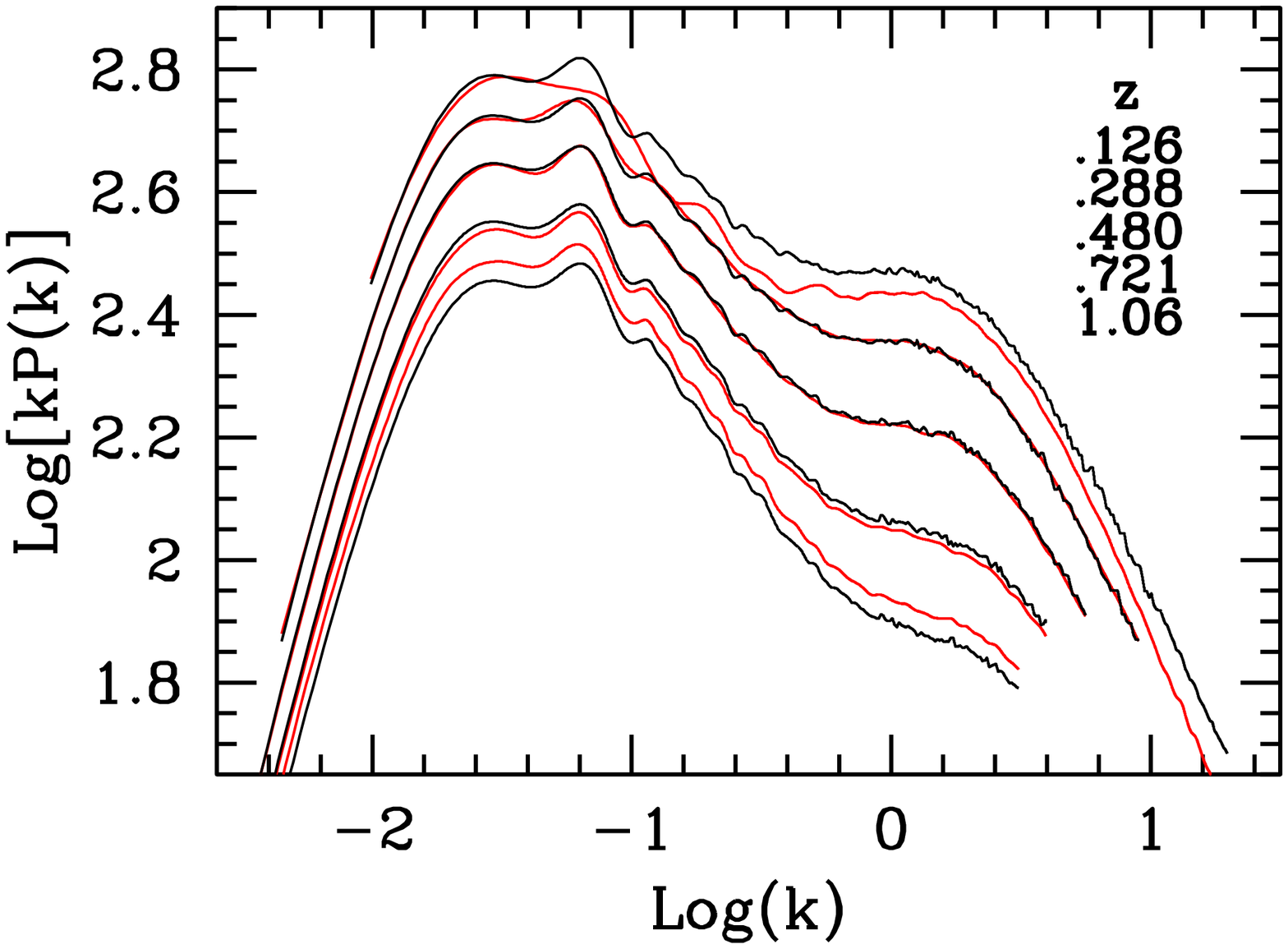}
%\vskip -2.truecm
\includegraphics[height=6.9cm,angle=0]{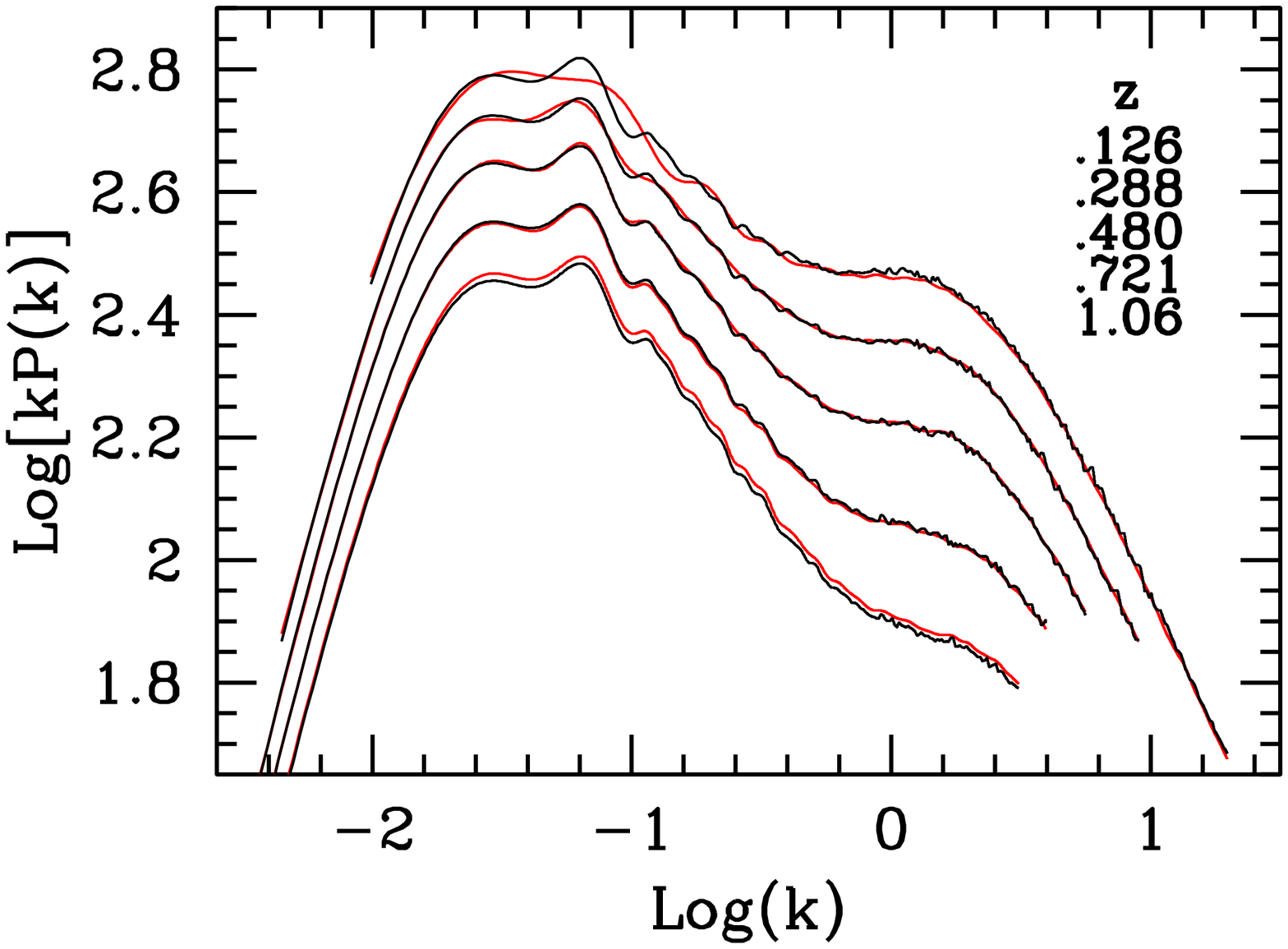}
\end{center}
\vskip -.6truecm
\caption{As previous Figure, just using RI spectra and 16 (l.h.s.) or
  15 (r.h.s.) non--zero $s_i$. }
\label{SPRI1516}
\end{figure}
%%%%%%%%%%%%%%%%%%%%%%%%%%%%%%%%%%%%%%%%%%%%%%%%%%%%%%%%%%%%%%%%%%%%%%%
%%%%%%%%%%%%%%%%%%%%%%%%%%%%%%%%%%%%%%%%%%%%%%%%%%%%%%%%%%%%%%%%%%%%%%%
\begin{figure}
\begin{center}
\vskip -2.2truecm
\includegraphics[height=6.9cm,angle=0]{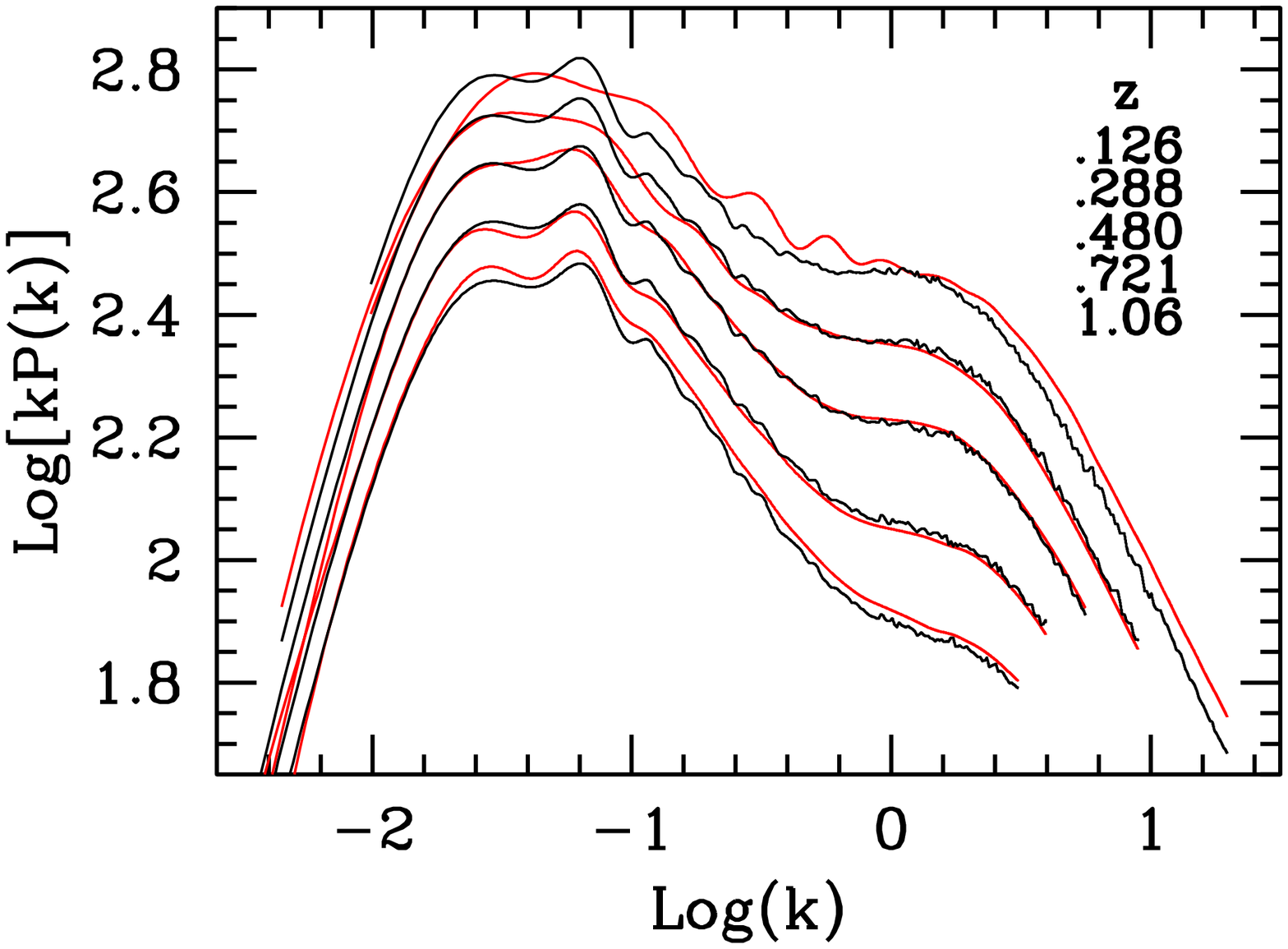}
\includegraphics[height=7.cm,angle=0]{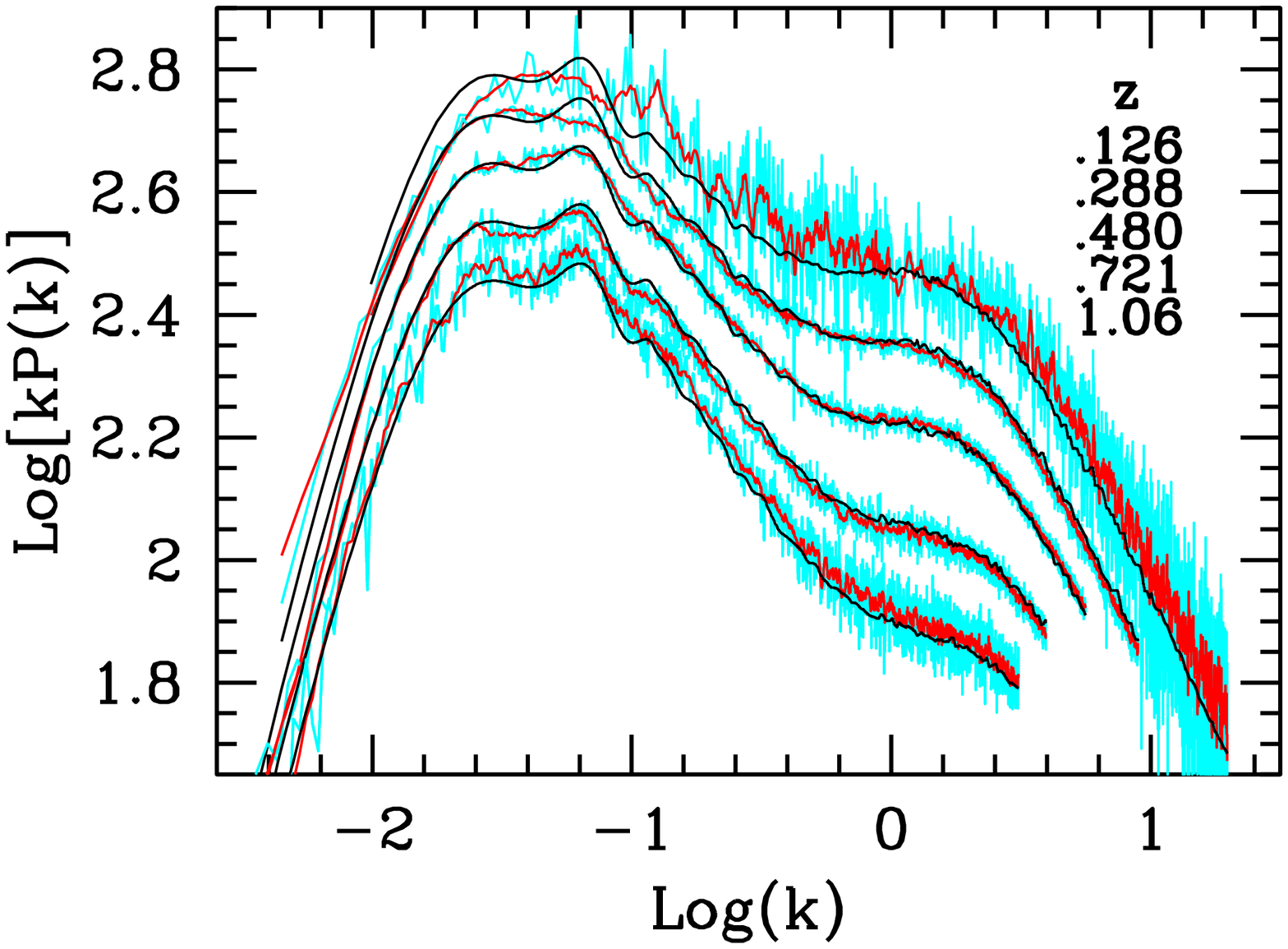}
\end{center}
\vskip -.6truecm
\caption{{\it L.h.s.:} As the previous Figure, just pushing the number
  of non--zero $s_i$ down to 6. {\it R.h.s.:} Inversion results, still
  with 6 non--zero $s_i$'s, after adding a noise $\simeq 0.8\,
  \%$. Mere inversion yields the cyan noisy curves. By applying a
  simple top--hat filter to them (average among 10 nearby $k$'s),
  errors are smoothed and we obtain the red curves.}
\vskip -.3truecm
\label{n0RI06}
\end{figure}
%%%%%%%%%%%%%%%%%%%%%%%%%%%%%%%%%%%%%%%%%%%%%%%%%%%%%%%%%%%%%%%%%%%%%%%
The passage from 8 to 7 o.o.m. in the $s_i$ selection is critical. In
the 21--$s_i$ case, we see the low--$z$ spectra covered by the
numerical noise. Notice also that the algorithm has however some
difficulty to recover the BAO shape, namely at low $z$.

In Figure \ref{SPGL2021} and in the following ones we plot a selection
of spectra, for the the same $z_i$ in all Figures, shown in the
frame. In addition to the values plotted, there are $\sim 20$
intermediate $z_i$ values for which spectra are recovered with a
precision similar to the best results shown.

We then apply the inversion algorithm, based on GL integration, on RI
spectra. The transition from biased to fair spectra then occurs at the
passage from 16 to 15 $s_i$ values. Unsatisfatory inversion is however
appreciable here through curve confusion, rather than by curves
obscured by numerical noise, as in Figure \ref{SPGL2021}. The
inversion is slightly less precise, typical errors being $\cal O$$(1\,
\%)$, apart of the BAO range, where recovery appears more
critical.

Before concluding the discussion on inversion in the absence of noise,
let us consider~al\-so the option of keeping just 6 $s_i$ (Figure
\ref{n0RI06}, l.h.s.). The efficiency of the SVD technique~is~so good
that it still allows to recover a fair deal of spectra: in the
$z$--range 0.2--1, $P(k,z)$ spectra are still recovered with errors
$\cal O$$(2\, \%)$ for most of the $k$ range; BAO's, however, tend to
be badly recovered and spectra normalization are close to fair only
for $0.2<z<1~.$

\subsection{Recovery in the presence of noise}
Error propagation from $C_{ij}(\ell)$ to $P(k,z)$ spectra is one of
the contributions of this work. As above mentioned, in the process of
matrix inversion errors expand. However, by suitably modeling
available parameters, error increase can be kept under control. The
cases shown here should be considered just as examples. Further
improvement is still possible, e.g. by pas\-sing to a more efficient
mapping from $u$ and $x$ variables. The noise in recovered $P(k,z)$
spectra is also filtered here in a simple fashion. More sophisticated
filtering is possible.

We start then from RI $C_{ij}(\ell)$ spectra. By using a Gaussian
random variable $G$, distributed around zero, whose variance is then
set to $\epsilon$, we build the ``noisy'' spectrum
\begin{equation}
C^{(N)}_{ij}(\ell) = C_{ij}(\ell) \times [1 +\varepsilon \, \, G]\, \, .
\label{noise}
\end{equation}
Cyan bands in Figure \ref{n0RI06} (r.h.s.) show inversion results,
obtained by keeping 6 non--zero $s_i$, with $\varepsilon = 0.8\, \%$.
Using 6 $s_i$'s yields a maximum/minimum $s_i$ ratio $\cal O$$(\sim
300)$ and this is also the theoretical upper limit to error
magnification.  Results are then filtered by using a 10--point
top--hat filter (red curves). In the central $z$--range, after
filtering errors are $\sim 5\, \%$, therefore keeping error
magnification well within a factor $\sim 10\, .$

The point here is that reducing the maximum/minimum $s_i$ ratio has a
twofold effect: it cuts off errors, but also kills the information in
the $\cal M$ matrix. It may then be useful to consider the results
shown in the Figures 8 in detail. The upper plots show the ratios
between recovered and input $P(k,z)$ for 12~equi\-spaced $z$ values,
when SVD inversion is performed by using 7, 6 (and 5) non--zero $s_i$
and 10--point top--hat smoothing is performed. In each frame, the 3
colors yield 3 redshift values (shown aside). Each curve can
be~as\-sociated to its redshift by reminding that a $z$ increase
shifts the $k$--range to the~left.

We are also confident that more sophisticated filters could reduce
errors by an extra factor $\sim 1/2$. If so, the expected error
magnification factor $\sim 300$ could be curbed down to $\sim 3$.

%%%%%%%%%%%%%%%%%%%%%%%%%%%%%%%%%%%%%%%%%%%%%%%%%%%%%%%%%%%%%%%%%%%%%%%
\begin{figure}
\begin{center}
\vskip -.5truecm
\includegraphics[height=7.cm,angle=0]{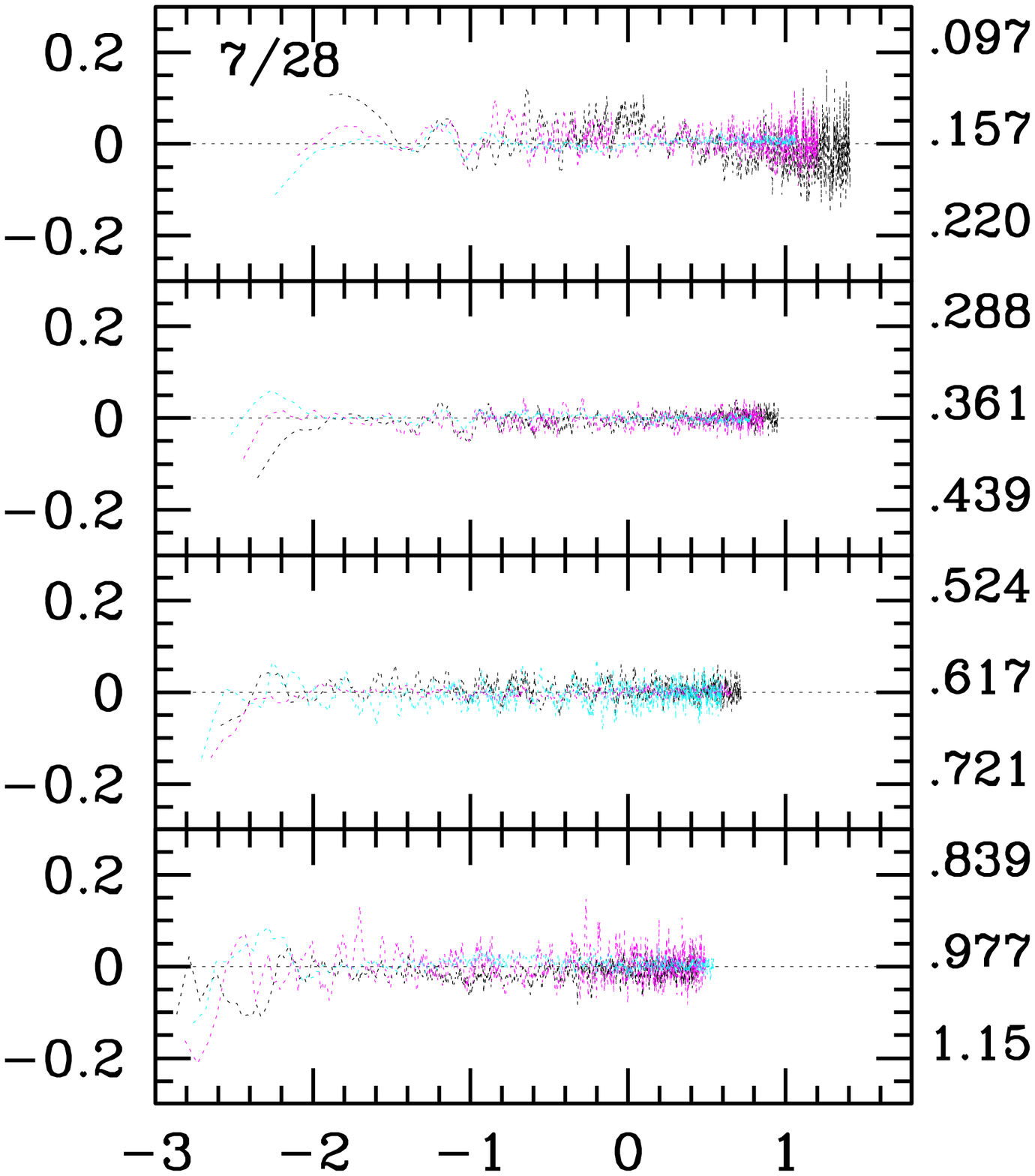}
\includegraphics[height=7.cm,angle=0]{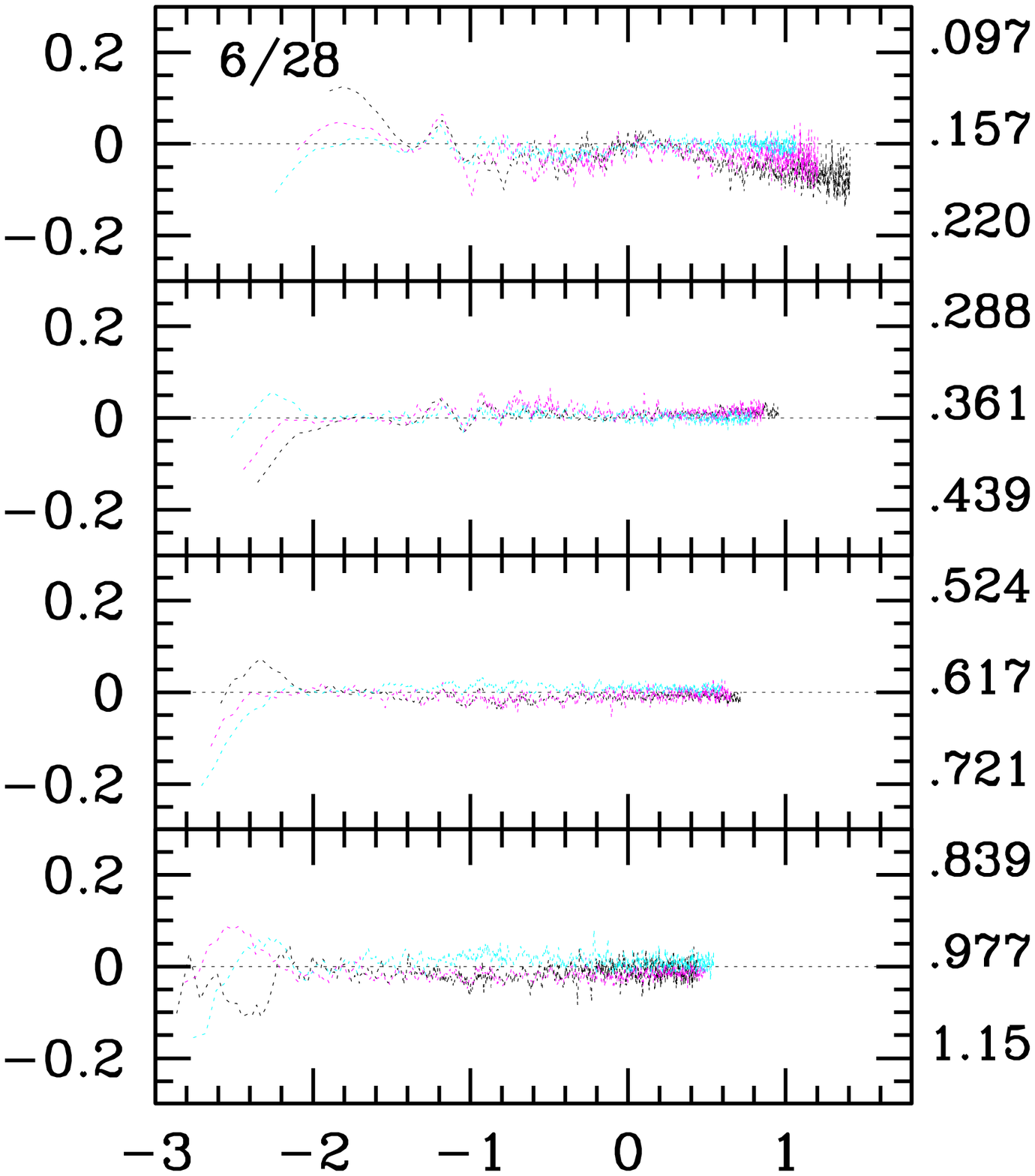}
\vskip -1.truecm
\includegraphics[height=6.9cm,angle=0]{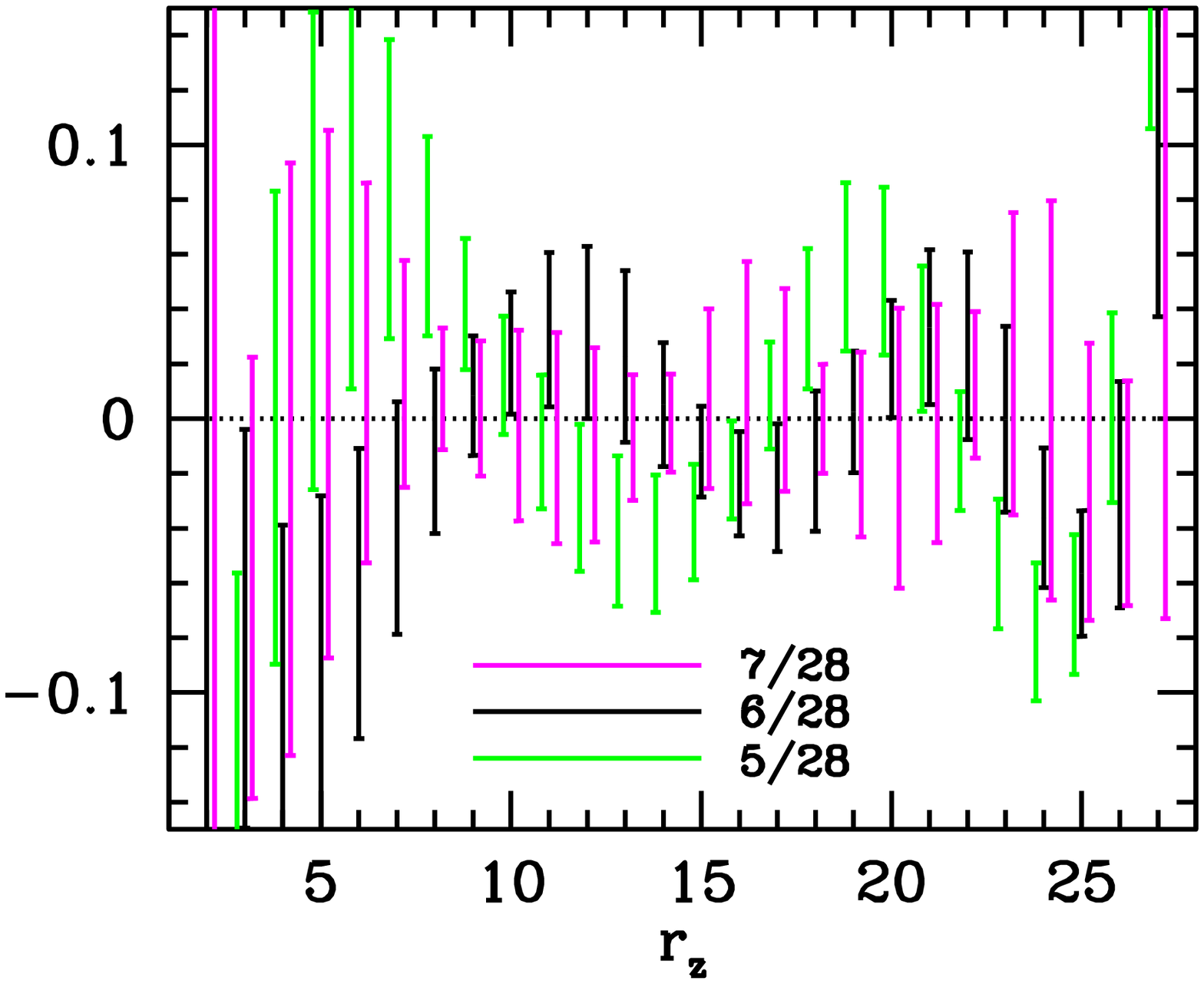}
\vskip -.5truecm
\caption{Ratio between recovered and input power spectra at different
  redshifts (at the r.h.s. of top Figures), when the shear spectra is
  added a Gaussian noise $\sim 0.8\,\%$. The recovered spectra where
  filtered through a 10-point top--hat smoothing. These plots are
  obtained when 7 or 6 non--zero $s_i$ are kept. Reducing the number
  of non--zero $s_i$ reduces error magnification, but biases the
  inversion process. In the bottom Figure average displacements from
  unity of the ratios between input and recovered spectra are shown;
  errorbars yield m.s. oscillation amplitudes around average. Best
  results seem obtainable with 6 non--zero $s_i$. }
\vskip -3.truecm
\end{center}
\vskip -.8truecm
\label{f11}
\end{figure}
%%%%%%%%%%%%%%%%%%%%%%%%%%%%%%%%%%%%%%%%%%%%%%%%%%%%%%%%%%%%%%%%%%%%%%%
The bottom frame however shown that reducing the number of non--zero
$s_i$'s yields a sort of oscillatory residuals, reminiscent of the use
of a finite number of orthogonal polynomials. Reducing to 5 the number
of non--zero $s_i$'s risks then to seriously bias the calibration of
fluctuation growth through redshifts. Probably, to keep on the safe
side, using 7 non--zero $s_i$ could be a fair choice. Testing advanced
error filters with realistic error distributions, for the 7--$s_i$
case, could tell us a decisive word on the precision achievable.

\section{Discussion}
In this paper we discuss the invertion of the integral expression
yielding tomographic shear spectra $C_{ij}(\ell)$ from density
fluctuation spectra $P(k,z)$. In Paper I, the question had been
discussed with (N=)5 bin tomography, yielding N(N+1)/2=15 independent
shear spectra. We however expect future data ({\sc Euclid}) to cover a
larger sky portion, so allowing up to 10 bin tomography. Here,
however, we considered just N=7, yielding N(N+1)/2=28 independent
shear spectra, however allowing us great improvements in respect to
N=5. Another contribution of this work is the study of noise
propagation from tomographic shear data to fluctuation spectra.

We have started from the equation yielding shear from fluctuation
spectra in the form
\begin{equation}
\label{cp}
c_A = {\cal M}_{A,r} p_r \, \, .
\end{equation}
Here $C_{ij}(\ell) \equiv H^4 c_A(\ell)$ and $p_r(\ell)=P(\ell/u_r,
u_r)$, $u$ being the comoving distance of the redshift $z$.  Eq.$\, \,
$(\ref{cp}) deals with each $\ell$ separately; summation replaces
there integration and is performed by using a Gauss--Laguerre (GL)
technique, by replacing the comoving distance $u$ by $ x = (u / \bar
u)^\beta $ and assuming a large--$u$ behavior $\propto \exp(-x)$.

As $A$ takes 28 values, we are allowed a 28--point GL integration; the
sum (\ref{cp}) then approaches the exact integral, even better. To
obtain the matrix $\cal M$ we need information on the space geometry,
as well as on the expected galaxy distribution on $z$ and the
statistical relation between photometric and spectroscopic $z$'s.

Recovering $P(\ell/u_r,u_r)$ from $C_{ij}(\ell)$ then requires
inverting the matrix $\cal M$. Unfortunately, as different bins refer
to different depths, $\cal M$ approaches a singular behavior.

The recovery of $P(k,z)$ is therefore based on the singular value
decomposition (SVD) technique, soon telling us the degree of
singularity through the elements of a 28$\otimes$28 diagonal matrix
$s$ derived from $\cal M$. The singularity however depends on the
choice of $\bar u$ and $\beta$ and, in this work, we provide criteria
for their selection.  {\bf In spite of optimization, however, the
  $SVD$ approach tells us soon that the $s_i$ components span $\sim
  13$ o.o.m.$\, $ (with the values adopted here, the maximum/minimum
  ratio is $\sim 1.22 \times 10^{13}$). Let us also outline that
  passing to $N>7$ gives $\cal M$ matrices yielding $s_i$'s spanning
  even more o.o.m.'s.; e.g., for $N=8$ the $s_i$ elements span 18
  o.o.m., at least.}

The SVD technique however allows us to reduce the effective divergence
level, by setting to zero a number of $s_i$. This is similar --$\,
$but not equivalent and, indeed, not so drastic$\, $-- to suppressing
a number of equations. Our analysis exploits this option, also because
error propagation depends on the ratio between minimum and maximum
non--zero $s_i$.

The simplest task that the inversion procedure can accomplish is the
recovery of $P(k,z)$ from $C_{ij}(\ell)$ obtainable via GL
integration. The only limitation that we meet, in this case, is the
divergence degree outlined by the diagonal $s$ matrix components. Out
of 28 $s_i$ components, we find that best results are obtainable when
keeping 20 non--zero $s_i$. Already with 21 components, numerical
noise affects the low--$z$ recovered spectral components.

We then tested the inversion of the results of exact integration.
Although being closer to realistic, this works as though adding a
noise to the $c_A(\ell)$ components. As a matter of fact, to obtain
fair results, we must then furtherly reduce the number of non--zero
$s_i$. Best results are obtained with 15 $s_i$ components. Typical
discrepancies between input and output fluctuation spectra are then
$\cal O$$(2\, \%)$, apart of some wider discrepancy arising in the
low--$k$ BAO range, namely for low redshift. Incidentally, this is a
great improvement in respect to 5 bins. Then, the very normalization
of recovered spectra was at risk and fair results could be obtained
only through renormalization in the low--$k$ range.

We then tested error propagation by superimposing a Gaussian noise
(m.s.a. $\sim 0.8\, \%$) onto the $C_{ij}(\ell)$ spectra obtained from
the integration of simulation spectra. Noise is magnified by matrix
inversion. A theoretical upper limit on noise magnification is set by
the ratio between the top and bottom $s_i$ kept (see Appendix B). This
calls for pushing further down the number of non--zero $s_i$. As a
matter of fact, best results are obtained when keeping 6 or 7
non--zero $s_i$ (over 28!). 

The noise in direct inversion results was then tentatively filtered by
using a 10--point top--hat smoothing. After using this admittedly
rough smoothing technique, we find that the overall noise
magnification is less than a factor 10, in respect of a theoretical
upper limit $\sim 300\, .$ Let us also outline that spectral values on
nearby $k$ values derive from independent $\ell$ inputs, so that
filtering acts in an ``orthogonal'' direction, in respect to matrix
inversion.

The $k$--region most penalized by the suppression of a large number of
$s_i$ is the BAO range. The first victim of the cut of $s_i$ is the
possibility to recover the BAO structure.

{\bf Finally, let us briefly outline the problems arising with $N>7$,
  because of the greater range of o.o.m.'s spanned by the $s_i$
  elements. With $N=7$ bands and in the most favorable case we have to
  suppress the matrix information related to 8 $s_i$, i.e. $\sim 30\,
  \%$ of it. Residual $s_i$ are within $\sim 7$ o.o.m.'s, allowing
  numerical routines $\sim 5$--6 o.o.m.'s, to work out reliable
  results.

In the same case, if $N=8$, to keep non--zero $s_i$ within 7 o.o.m.'s,
one must give up almost half of them. Spectral recovery still works,
but errors are wider. The problem is clearly in numerical precision.
Using quadruple precision in not trivial, but possible, and would
allow us to exploit $N>7$ bands.  }

\section{Conclusions}
Shear data are expected to be a new and effective resource to
discriminate among cosmological models. They are most sensitive to the
rate of expansion and fluctuation growth at low--$z$, when the
contribution of Dark Energy is determinant. Accordingly, they are
expected to be an outstanding probe on the DE equation of state $w(z)$
and on possible deviation from GR, or to couplings between the dark
cosmic components.

% ellipse (on)

From data we expect tomographic shear spectra to be derived. A basic
difficulty amounts to cleansing rough data from intrinsic
ellipticities. Here we assume that this aim has been achieved and that
we can use any number of bins up to 10. 

An analysis of spectral data can be performed in accordance with a
Bayesian paradigm: $C_{ij}(\ell)$ model predictions are then made and
compared with $C_{ij}(\ell)$ data and their errorbars.

% ellipse (off)

%The basic analysis of spectral data will be performed in accordance
%with a Bayesian paradigm.  In this context, $C_{ij}(\ell)$ model
%predictions will be made, and compared with $C_{ij}(\ell)$ data and
%their errorbars.

Other datasets, also from the same {\sc Euclid} experiment, will
directly concern $P(k,z)$ spectra. The option to skip $C_{ij}(\ell)$
model predictions, by directly~associ\-ating $P(k,z)$ spectra derived
from them with other $P(k,z)$ data, is then appealing. By using double
precision routines, best results are obtainable if $C_{ij}(\ell)$ come
from 7 bins. A rather satisfactory test on error magnification in this
derivation has also been performed here, but it is perhaps premature
to state that errors are under control. This seems the main point to
be further deepened in future work.

To build the $\cal M$ matrix, whose inversion allows us
% ellipse (on)
to work out $P(k,z)$ from $C_{ij}(\ell)$,
% ellipse (off) 
we need to convert redshifts into distances. Model {\it geometry} must
then be known, but neither $\sigma_8\, \, $ and $n_s$, nor parameters
describing baryon physics (including $\Omega_b$) shall be input.

For the sake of example, let us then outline a possible investigation
pattern: SNIa data could become so good that the distance modulus
could provide $u(z)$ with negligible errors. From such $u(z)$ the
parameters of background cosmology can be derived. However,
independently from such derivation, we can use $u(z)$ to build the
window functions $W_r(u)$ and derive from them the inverted matrix ${\cal
  M}^{-1}$,
% ellipse (on) 
so obtaining $P(k,z)$ --$\, $apart of a constant factor $(\Omega_m
H_0^2)^2\, $--
% As a matter of fact, to do so, we still lack
% the knowledge of $\Omega_m$; therefore, ${\cal M}$ and its inverse are
% $ nown apart of a constant factor $\Omega_m^2$. By applying it to shear
% data, we could then work out $P(k,z)$ apart of a constant factor
% $(\Omega_m H_0^2)^2$. This, however, would not inhibit us from
and the growth law $G(a)$. In parallel with fitting $u(z)$
to models, we then also fit $G(a)$ to them, so performing two
independent model tests, based on geometry and dynamics, respectively.
The two tests could provide a direct confirm/falsification of models
where DE and DM are two real and independent cosmic components.

More in general, this technique allows us to measure the fluctuation
growth in an unbiased way, over any $k$--range, namely those where
different hypotheses on stellar formation, SN explosion, AGN energy
release, etc., are critical. Accordingly, the inversion technique
might become an important tool to discriminate among different
hypotheses on baryon physics, possibly testing parameters beyond those
spanning the bayesian parameter space.

Let us then outline that there is still room for further improvements
of the technique providing $P(k,z)$ from shear spectra. In particular,
let us stress the following points: (i) A great improvement was
achieved when passing from N=5 to N=7; further improvements may be at
hand if using 8, 9, or 10 bins.  This just requires going beyond
double precision.
% ellipse (on)
%Although {\sc Euclid} is expected to provide a tomography with up to
%10 bins, it is not clear to us up to which point intrinsic shear
%cleansing is more effective with $N = 5$ or less.
% ellipse (off) 
(ii) Noise propagation was considered here to test error propagation.
Propagated noise was then filtered, with an effective but rought
procedure. Better filtering techniques are surely available, but
selecting among them depends on the nature of the ``noise'' used to
describe errors.  (iii) A technical issue concerns the passage from
the coordinate $u$ to the integration variable $x$. The choise of
$\bar u$ and $\beta$ in the conversion expression $x=(u/\bar u)^\beta$
was found to be critical. No $u$--$x$ mapping more evolute than a
simple power low was however considered; their use in the low--$x$
range could allow us further improvements.

Since a few decades, various authors devoted their efforts to invert
the {\it Limber equation} \cite{Limber}, aiming at obtaining the
spatial correlation function from the angular one, without making an
{\it ansatz} on its shape. Nice algorithms were built, also taking
into account relativistic effects \cite{GPP,BL,ez}. Unfortunately,
none such algorithm really found a practical application to data
analysis.

Also the equations inverted here are a sort of Limber equation, but
one should avoid to infer, from past failures, that this approach will
face unsurmontable practical difficulties, as well. Of course, {\it
  lensing} angular spectra, instead of angular positions, are used
here and, in a sense, this makes the problem harder. However, {\it
  tomographic} data are used here: a third coordinate is therefore
input and the whole technique makes an extensive use of it.

\vskip .6truecm

\noindent
ACKNOWLEDGMENTS. SAB acknowledges the support of CIFS. We are grateful
to Stefano Borgani for making available to us his large hydrodynamical
simulations and to Volker Springel for the non-public GADGET-3 code
used to run them. Giuseppe La Vacca is also gratefully thanked for a
number of useful discussions. H.~Hoekstra, T.~Kitching, L.~Guzzo and
W.~Percival are also to be thanked for comments.

%\vskip .3truecm
%\vskip .3truecm

% Bibliography

%\vskip 1.truecm

\centerline{\bf Appendix A}

\vglue .4truecm
\noindent

\rm
\vskip .01truecm

\noindent
{\bf 1. A brief description of the simulation used}

\noindent
The simulation box side $L = 410\, h^{-1}$Mpc ($k_L=2\pi/L \sim 1.5
\times 10^{-2}\, h\, $Mpc$^{-1}$); $(2 \times) 1024^3$ particles are
used, with a force resolution $\epsilon = 7.5 \, h^{-1}$ kpc
($k_\epsilon = 2\pi/\epsilon \sim 8.4 \times 10^{2}\, h$Mpc$^{-1}$).
Spectra can be used up to $k \simeq N(2\pi /L)$ with $N \simeq 2^{15}
= 32768$, however keeping $k~( \simeq 5.0 \times 10^{2}\, h$Mpc$^{-1})
< k_\epsilon$, provided that they are not covered by shot noise (see
\cite{borgani}).

According to Table I of Section 2, CDM and baryon particle masses are
$m_c \simeq 1.89 \times 10^9\, h^{-1} M_\odot$ and $m_b \simeq 3.93
\times 10^8\, h^{-1} M_\odot$, respectively. Baryon dynamics includes
cooling and star formation. Metal production from chemical enrichment
contributed by SN-II, SN-Ia and AGB stars, as described in
\cite{tornatore}, is also included. Stars, distributed according to a
Salpeter IMF, release metals over the time-scale determined by the
corresponding mass-dependent life-times. Kinetic feedback is also
implemented by mimicking galactic~ej\-ecta powered by SN explosions. AGN
feedback is not included, at variance from \cite{sembol}, who however
used smaller boxes. The simulation start at $z=41$ and spectra are
obtained at
\vskip -.4truecm
\begin{equation}
1+z_r = 10^{r/20} \, \, \, \, \, \, \, (r=1,\, ....\, ,19)\, \, .
\end{equation}
\vskip -.2truecm
\noindent
8 of them, up to $z \simeq 1.24$ are between 0 and 1.4; 7 more are
within $z=4.02$. For more details see \cite{borgani}. These 15 spectra
are used for interpolation.

\vskip .4truecm

\noindent
{\bf 2. Spectra interpolation}

\begin{figure}
\begin{center}
\vskip -.5truecm
\includegraphics[height=6.7cm,angle=0]{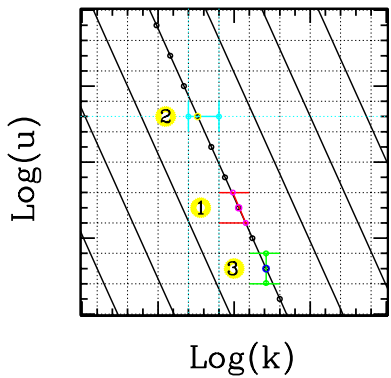}
\end{center}
\vskip -1.1truecm
\caption{Interpolation patterns. In the Figure we assume ``data'' to
  be available at the crossings between horizontal and vertical dotted
  lines. The tilted lines are at constant $\ell$.  Along one of them,
  we indicate a typical set of points where we may need to know
  $P(l/u,u)$, to perform the numerical integration. To reach them, in
  any case, there are 2 interpolation steps. The direction of former
  (latter) is indicated by the two parallel segments (a single
  segment, tilted or orthogonal).  }
\label{trpl}
\end{figure}

\noindent
There are at least 3 different patterns to interpolate, schematically
shown in Figure \ref{trpl}.  
%%%%%%%%%%%%%%%%%%%%%%%%%%%%%%%%%%%%%%%%%%%%%%%%%%%%%%%%%%%%%%%%%%%%%%%
%%%%%%%%%%%%%%%%%%%%%%%%%%%%%%%%%%%%%%%%%%%%%%%%%%%%%%%%%%%%%%%%%%%%%%%

\noindent
(1) We can first work out $P_\ell(u_a) \equiv
P(\ell/u_\alpha,u_\alpha)$ at fixed $\ell$, by interpolating among
$k_b$ values. To perform the integral we need $P_\ell(u)$ at suitable
values of $u \neq u_a$, and this requires a further interpolation at
constant $\ell$. (2) We can first determine $P(k_b,u)$ at the $u$
value needed; then interpolate at constant $u$ along $k$.  (3) We can
first determine $P(k,u_a)$ at the $k$ value needed; then we
interpolate at constant $k$ along $u$.

Interpolation involves more than the 2 closest points in each
direction. Thick segments connecting 2 close points are just as an
indication of the directions along which one uses a cubic spline.

By using the {\sc Halofit} algorithm, yielding spectra at arbitrary
$k$ and $u$ we found that the procedures (2) \& (3) perform
similarly. On the contrary, the procedure (1) (interpolation along
tilted $\ell$--constant curves) yields consistent results only when
spectra are known at $\sim 3$ times more redshift values $z_a$.  Still
using {\sc HALOFIT} we tested that our simulation ``data'' is adequate
for procedures (2) and (3), as far as $k$ and $z$ values available.

\vskip .5truecm

\centerline{\bf Appendix B}

\vglue .4truecm
\noindent
{\bf 1. The singular value decomposition (SVD)}

\rm
\vskip .2truecm

The SVD technique is based on a theorem of linear algebra stating that
any $M \times N$ ($M > N$) matrix $\cal M$ can be decomposed, in a
unique way apart of multiplicative factors, into the matrix product $U
\cdot s \cdot V^T$ ($V^T$ is the transpose of a matrix $V$). More in detail:
$$
\left|\matrix{ {\cal M}_{11} & ... & {\cal M}_{N1} \cr
         ....  & ... & ....  \cr
         {\cal M}_{1M} & ... & {\cal M}_{NM} } \right|
=
\left|\matrix{ U_{11} & ... & U_{N1} \cr
         ....  & ... & ....  \cr
         U_{M1} & ... & U_{NM} } \right| \cdot
\left|\matrix{ s_1 & 0  & 0 \cr
              ... & ... & ... \cr
              0 & 0 & s_N } \right|
\cdot
\left|\matrix{ V^T_{11} & ... & V^T_{N1} \cr
         ....  & ... & ....  \cr
         V^T_{N1} & ... & V^T_{NN} } \right| ~,
\eqno (B1)
$$ 
with a diagonal $s$, while $U$ and $V$ are orthogonal, i.e.,
$$
\sum_i U_{ia} U_{ib} = \delta_{ab}~,~~~ 
\sum_i V_{ia} V_{ib} = \delta_{ab}~,
\eqno (B2)
$$
or
$$
U^T \cdot U = {\mathbf 1} ~,~~~ V^T \cdot V = {\mathbf 1} ~.
\eqno (B3)
$$ 
In this work we are interested just in the case when $M=N$ and all
matrices are square; accordingly, it is also $ U \cdot U^T = {\mathbf
  1}$ (and $ V \cdot V^T = {\mathbf 1}) $.  The {\it condition number}
of the matrix $\cal M$ is then defined as the ratio $K$ between the
largest and smallest (in magnitude) $s_i$. A matrix is said to be {\it
  ill--conditioned} if $K$ is too large (or {\it singular} if $K \to
\infty$).

\vskip .5truecm
\noindent
{\bf 2. Norms and error magnification}

\vglue .2truecm 

Let us recall first that the (2--)norm of a vector {\bf x} reads
$$
|{\bf x}| = \left[ \sum_i |x_i|^2 \right]^{1/2}
$$
$x_i$ being its components. The (2--)norm of a matrix {\bf A}
is then defined as follows:
$$
|{\bf A}| = max \left( |{\bf A \cdot x}| \over 
|{\bf x}| \right)
\eqno (B4)
$$ 
i.e.: $|{\bf A}|$ is the maximum factor by which the matrix {\bf A}
can amplify a non--zero vector {\bf x}. It can then be shown that, if
we perform a SVD of {\bf A} and $s_1$ is the top component of the
diagonal matrix ${\bf s}$, in the relation $ {\bf A} = {\bf U} \cdot
{\bf s} \cdot {\bf V}^{-1}~, $ it is
$$
|{\bf A} | = s_1~.
$$
Let us then outline that, if we state $t_i = s_i^{-1}$, it is
$
{\bf A}^{-1} = {\bf V} \cdot {\bf t} \cdot {\bf U}^{-1}~.
$ 
Accordingly, $ |{\bf A}^{-1}| = 1/s_n~, $ $s_n$
being the smallest component of the matrix ${\bf s}$ and
$$
|{\bf A} | \cdot |{\bf A}^{-1} | = s_1/ s_n~.
\eqno (B5)
$$
The matrix--vector and matrix--matrix products 
have then the following properties:
$$
|{\bf A \cdot x}| \leq |{\bf A}| \cdot|{\bf x}|
~~~~{\rm and} ~~~~
|{\bf A \cdot B}| \leq |{\bf A}| \cdot|{\bf B}| ~.
\eqno (B6)
$$

Let us then assume that
$$
{\bf A} \cdot {\bf x} = {\bf b}
\eqno (B7)
$$ 
and consider $\tilde {\bf x} = {\bf x} + \delta {\bf x}$, as we do
when we try to reobtain $P(k,z)$ from noisy $C_{ij}(\ell)$. It shall
be $ {\bf A} \cdot \tilde {\bf x} = {\bf b} + \delta {\bf b} $ and
therefore
$
\tilde {\bf x} - {\bf x} = {\bf A}^{-1} \cdot \delta {\bf b}~,
$
so that, owing to eq.~(B6), 
$
|\delta {\bf x}| \leq |{\bf A}^{-1}| \cdot |\delta {\bf b}|~.
$
In turn, eq.~(B7) yields
$$
{1 \over |{\bf x}|} \leq {|{\bf A}| \over |{\bf b}|}
~~~~{\rm 
and,~ therefore
}~~~~
{|\delta {\bf x} | \over |{\bf x}|}
 \leq |{\bf A}| \cdot|{\bf A}^{-1}| \cdot {|\delta {\bf b}| \over |{\bf b}|}~.
$$
Owing to eq.~(B5) we have then that
$$
{|\delta {\bf x} | \over |{\bf x}|}
 \leq {s_1 \over s_n} \cdot {|\delta {\bf b}| \over |{\bf b}|}~.
\eqno (B7)
$$ 
i.e., that error magnification is surely smaller than the ratio
between the maximum and minimum $s$ components considered, coinciding
with the {\it condition number} $K$.

For the proofs of the theorems, besides of \cite{NR,golub}, see also
\cite{pttb}.

\vskip .4truecm
\noindent
{\bf 3. Gauss--Laguerre parameter selection}

\vglue .2truecm

The Gauss--Laguerre sum, replacing the integration over $P(k,z)$ to
obtain the shear spectra $C_{ij}(\ell)$, ought to by made with an eye
to the matrix inversion which will follow. The condition number $K$ of
the matrix $\cal M$, in fact, depends in a critical way on the
selection of the $\bar u$ and $\beta$ parameters used in the change of
variable $ x = (u/\bar u)^\beta~.$ A bad selection of $\bar u$ and
$\beta$ can yield $K$ values up to $10^{20}$ or even larger. Moreover,
$K$ is also directly linked to error magnification, according to
eq.~(B7).

To minimize $K$ we can act on 3 parameters: $\beta$, $\bar{u}$, and
$N_{cut}$, the number of $s_i$ which (unavoidably) will be set to
zero.  Increasing $N_{cut}$ obviously reduces $K$, but our aim amounts
to minimizing $K$ while keeping $N_{cut}$ sufficiently low, so to
grant a fair and physical solutions to the a substantial part of the
28 equations forming the linear system.

Let us then keep into account that the galaxy redshift distribution
$n(z)$ rapidly declines beyond $z \sim 1$ and that, therefore, the
window functions $W_r$ and the $\cal M$ matrix element also fade. In
the following we shall not directly consider contributions coming from
above a maximum redshift $z_{max}=1.4$, corresponding to a maximum
comoving distance $u_{max} = 4194\, $Mpc, a choice yielding fair
results. We shall then select the value of $\bar{u}$ so that
\vskip -.4truecm
$$
  \bar{u}(\beta) = {u_{max}}/{x_{M}^{1/\beta}}
$$
\vskip -.02truecm
\noindent
$x_{M}$ being the largest node in the Gauss-Laguerre integration.

With this rule, we can then estimate the condition number for a wide
range of $\beta$ (from $0.1$ to $3.5$) and $N_{cut}$ values.  The
latter parameter will be considered in the interval 0 (all singular
values kept) --26 (only the two largest singular values kept, to
estimate their ratio).

\begin{figure}
\begin{center}
\vskip -.5truecm
\includegraphics[height=9.5cm,angle=0]{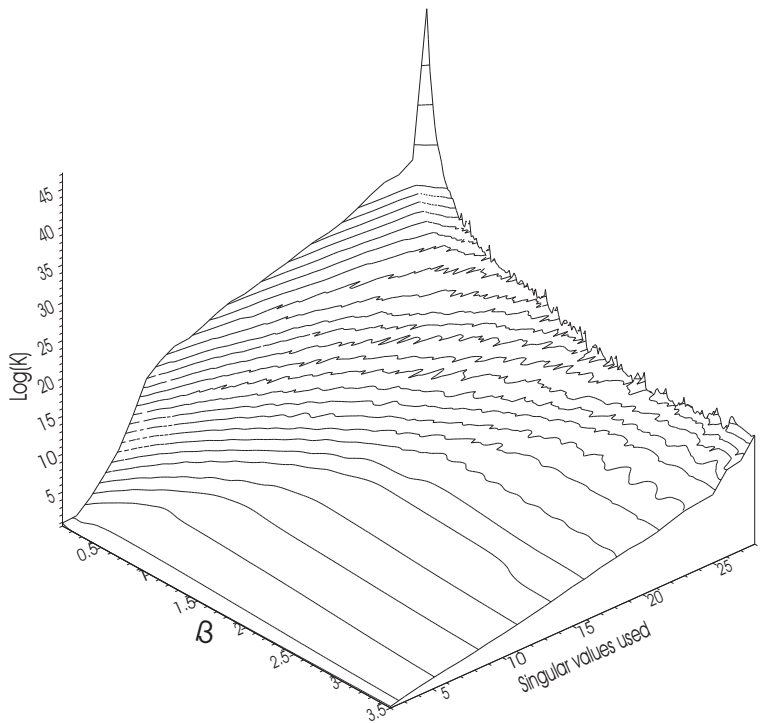}
\end{center}
\vskip -.3truecm
\caption{The condition number $K$ {\it vs.} the number of non--zero
  $s_i$ kept and $\beta$. At high $K$, there appear some graphic
  irregularities in the curves, due to the discreteness of the $s_i$
  number. }
\label{wireframe}
\vskip -.4truecm
\end{figure}

In Figure \ref{wireframe}, $K$ is plotted as a function of $\beta$ and
of the residual $s_i$ number ($28-N_{cut}$). The plot shows how
dramatically $K$ blows up just by choosing unsuitable values of the
free parameters. It is also evident that, reducing the number of
non--zero $s_i$, $K$ sistematically decreases and that, for a given
number of non--zero $s_i$, $K$ depends on $\beta$.

However, when the number of non-zero $s_i$ is small ($< \sim 10$) the
$K$ dependence on $\beta$ weakens. This can be exploited to perform
slight changes of the $\beta$ value, yielding a suitably different set
of $u_r$ for the Gauss--Laguerre nodes $x_r$. In this way, we can
obtain $P(k)$ at almost any $z$ value, in the $z$--range where
matrix inversion is efficient (see text).

%%%%%%%%%%%%%%%%%%%%%%%%%%%%%%%%%%%%%%%%%%%%%%%%%%%%%%%%%%%%%%%%%%%%%%%
%%%%%%%%%%%%%%%%%%%%%%%%%%%%%%%%%%%%%%%%%%%%%%%%%%%%%%%%%%%%%%%%%%%%%%%

\end{document}